\documentclass[useAMS,usenatbib]{mn2e}

\usepackage{latexsym}
\usepackage{amsmath}
\usepackage{amssymb}
\usepackage{graphicx}
\title[On the impact of the magnitude of Interstellar pressure on physical properties of Molecular Cloud]{On the impact of the magnitude of Interstellar pressure on physical properties of Molecular Cloud}
\author[Anathpindika. S, Burkert. A and Kuiper, R.]{Anathpindika, S $^{1},^{2}$\thanks{E-mail: sumed$\_$k@yahoo.co.in (SVA)}, Burkert, A$^{3},^{4}$\thanks{Max-Planck Fellow} and Kuiper, R$^{2}$ \\
$^{1}$Indian Institute of Technology, Kharagpur, West Bengal, India\\
$^{2}$ Institute for Astronomy \& Astrophysics(IAAT), University of T$\ddot{\mathrm{u}}$bingen, 10 Auf Der MorgenStelle, T$\ddot{\mathrm{u}}$bingen, Germany\\
$^{3}$ University Observatory Muenich, Schneirstrasse 1, 81679, Muenich, Germany \\
$^{4}$ Max Planck Institut f$\ddot{u}$r extraterrestrische Physik, Geissenbachstrasse, 85748, Garching, Germany \\}
\begin{document}

\date{Accepted 0000 December 00. Received 0000 December 00; in original form 0000 October 00}

\pagerange{\pageref{firstpage}--\pageref{lastpage}} \pubyear{2002}

\maketitle

\label{firstpage}

\begin{abstract}
Recently reported variations in the typical physical properties of Galactic and extra-Galactic molecular clouds ({\small MC}s), and in their ability to form stars have been attributed to local variations in the magnitude of interstellar pressure. Inferences from these surveys have called into question two long-standing beliefs that the {\small MC}s : \textbf{(1)} are Virialised entities, and \textbf{(2)} have 
 approximately constant surface density i.e., the validity of the Larson's third law. In this work we invoke the framework of cloud-formation via collisions between warm gas flows.
Post-collision clouds forming in these realisations cool rapidly and evolve primarily via the interplay between the Non-linear Thin Shell Instability ({\small NTSI}), and the self-gravity. Over the course of these simulations we traced the temporal evolution of the surface density of the assembled clouds, the fraction of dense gas, the distribution of gas column density ({\small N-PDF}), and the Virial nature of the assembled clouds. We conclude, these physical properties of {\small MC}s not only exhibit temporal variation, but their respective peak-magnitude also increases in proportion with the magnitude of external pressure, $P_{ext}$. The velocity dispersion in assembled clouds appears to follow the power-law, $\sigma_{gas}\propto P_{ext}^{0.23}$. Also, the power-law tail at higher densities becomes shallower with increasing magnitude of external pressure, for magnitudes, $P_{ext}/k_{B}\lesssim 10^{7}$ K cm$^{-3}$; at higher magnitudes such as those typically found in the Galactic {\small CMZ} ($P_{ext}/k_{B} > 10^{7}$ K cm$^{-3}$), the power-law shows significant steepening. Thus while our results are broadly consistent with inferences from various recent observational surveys, it appears, {\small MC}s hardly exhibit a unique set of properties, but rather a wide variety, that can be reconciled with a range of magnitudes of pressure between 10$^{4}$ K cm$^{-3}$ - 10$^{8}$ K cm$^{-3}$.
\end{abstract}


\section{Introduction}
Recently reported results by Rice \emph{et al.} 2016 (hereafter, Rice \emph{et al.}), after a fresh analysis of the ${}^{12}$CO data from the CfA-Chile survey of the Galactic Giant molecular clouds (GMCs), lends further credence to the hypothesis that the prevalent ambient conditions in the Galactic disk likely influence the physical properties of GMCs. As in an earlier study of physical properties of GMCs in the Galactic Ring Survey by Heyer \emph{et al.} (2009), Rice \emph{et al.} also report a variation in the coefficient associated with the Larson's first law (Larson 1981), the so-called size-linewidth scaling relation ($\sigma\equiv b\mathcal{L}^{a}; a=0.38, b=1.1$), as a function of the position of a {\small GMC} in the Galactic disk.  In particular, Rice \emph{et al.} reported, $b\equiv 0.5\pm0.05$ for clouds located in the inner Galactic disk and $b\equiv 0.38\pm0.05$ for those located farther out; the exponent, $a$, for all the clouds in their sample was close to $\sim 0.5$. This observed variation in the magnitude of the coefficient, $b$, has been inferred as evidence suggesting a corresponding variation in the linewidth; the larger magnitude of $b$ for clouds located in the inner Galactic ring has been attributed to those clouds having a larger velocity dispersion. Heyer \emph{et al.} (2009), on the other hand, reported that $b\equiv\Big(\frac{\sigma}{\mathcal{L}^{a}}\Big)$, in fact varied in proportion with the square-root of the gas column density, $\Sigma_{gas}$. While the recent findings reported by Rice \emph{et al.}, like one of the earliest studies of Galactic clouds reported by Solomon \emph{et al.} (1987), still imply that clouds in the Galactic disk are approximately consistent with the Simple Virial equilibrium ($\sigma\propto\mathcal{L}^{0.5}$; {\small SVE}), their conclusion is inconsistent with that deduced by Heyer \emph{et al.} (2009).  The reason for this inconsistency is probably the difference in the tracer used by the respective authors : Rice \emph{et al.} used the more common ${}^{12}$CO emission line to detect clouds, on the contrary, Heyer \emph{et al.} (2009) used the isotopologue ${}^{13}$CO which is known to be a more reliable tracer at higher column densities. In spite of this difference, conclusions  from these two surveys unambiguously demonstrate the variation in the linewidth across clouds located in different regions of the Galactic disk. \\ \\
In other related work, Hughes \emph{et al.} (2010) reported a size-linewidth relation, $\sigma\equiv 0.18\mathcal{L}^{0.74}$, for clouds in the {\small LMC}. The obvious conclusion being that these clouds have significantly smaller line-widths as compared with the Galactic clouds. Furthermore, these authors also point to the fact that the H$_{2}$ mass surface density appears to increase with increasing magnitude of the interstellar pressure, $P_{ext}$.
These reported findings were further corroborated by Hughes \emph{et al.} (2013b) in their study of {\small GMC}s in other {\small MW}-like galaxies such as the {\small M51} and {\small M33}. Not only did Hughes \emph{et al.} (2013b) find variations in the H$_{2}$ surface density across clouds, but also reported variation of the coefficients $(a,b)$ in the size-linewidth relation; the inferred line-widths for {\small GMC}s in the {\small M51} were higher than those for clouds in the {\small M33} and the {\small LMC}. These results challenge our extant beliefs and  understanding about cloud-properties. Previous studies, for instance those by Rosolowsky \emph{et al.} (2003; 2007) and Bolatto \emph{et al.} (2008) argued that physical properties of Galactic clouds and indeed of those in other nearby galaxies are approximately uniform, or rather, universal. These conclusions were likely a consequence of {\small GMC} properties being derived only from ${}^{12}$CO observations and were therefore a mere reflection of the physical conditions necessary to produce this emission. These emission peaks only identified the high density regions that are likely to be immune to the environmental conditions.\\ \\
Observational evidence pointing to variations in {\small GMC} properties have been variously interpreted : \textbf{(i)} Dobbs \emph{et al.} (2011) argued that the deviation of observationally deduced size-linewidth relation from the canonical relation possibly implied that clouds are largely unbound entities, \textbf{(ii)} Heyer \emph{et al.} (2009) suggested that cloud masses in their survey could have been systematically underestimated leading to the reported deviation, and \textbf{(iii)} the reported data could have been influenced by the external pressure, $P_{ext}$, a rather old idea (e.g. Bonnor 1956, Keto \& Myers 1986, Elmegreen 1989, Bertoldi \& McKee 1992, Field \emph{et al.} 2011).  In fact, Elmegreen (1989), having argued that clouds were entities that probably marked an approximate equipartition between self-gravity, external pressure and their internal energy, deduced the relation
\begin{equation}
\sigma\propto\Big(\frac{P_{ext}/k_{B}}{10^{4} \mathrm{K\ cm^{-3}}}\Big)^{1/4}\Big(\frac{\mathcal{L}}{\mathrm{pc}}\Big)^{1/2}.
\end{equation}
Alternatively, Field \emph{et al.} (2011) proposed that the observed variations in the size-linewidth relation could be reconciled if the clouds obeyed the (external)pressure-modified Virial equilibrium ({\small PVE}), instead of the usual Simple Virial Equilibrium ({\small SVE}). The implication being, a single magnitude of the external pressure, $P_{ext}$, may be insufficient to explain the variations reported by Heyer \emph{et al.} (2009). Instead, their data could be reconciled only if different clouds in the sample experienced different magnitudes of pressure in the range 10$^{4}$ K cm$^{-3}$ - 10$^{7}$ K cm$^{-3}$. Equivalently, it meant, the size-linewidth relation would be modified by the magnitude of external pressure, $P_{ext}$, experienced by a {\small GMC}. In an earlier work, Ballesteros-Paredes (2006) strongly argued against {\small GMCs} obeying the {\small SVE} because : \textbf{(i)}  clouds being the consequence of turbulent fragmentation in the interstellar medium ({\small ISM}), the fragmentation process must necessarily induce a flux of mass, momentum and energy between clouds and the {\small ISM}, and \textbf{(ii)} clouds often exhibit asymmetries in their respective line-profiles which is inconsistent with clouds obeying the {\small SVE}. Instead, the Larson's scaling relations should, at best, be viewed as evidence for energy-equipartition. The possibility of bound, turbulence-supported clouds as suggested by Krumholz \& McKee (2005) is difficult to reconcile, for it is unlikely there will be no redistribution of mass within such clouds over their lifetime (Ballesteros-Paredes 2006). Gas dynamics within a typical {\small GMC} is likely governed by collisions  between smaller clumps within them (e.g. Anathpindika 2009 a,b and references there-in), and/or feedback from existing stellar populations (see e.g. MacLow \& Klessen 2004 and references there-in). \\ \\
In a more recent work, Ballesteros-Paredes \emph{et al.} (2011) argued that the variations in the size-linewidth relation reported by Heyer \emph{et al.} (2009) is actually consistent with the scenario of hierarchical fragmentation in which a cloud does not collapse as a whole, but only isolated pockets in it collapse to form stars. In this paradigm, although turbulence in the {\small ISM} possibly plays a role in assembling the cloud, post its formation, when the contribution due to self-gravity becomes significant, the cloud no longer need be in equilibrium with the external medium. It is therefore unnecessary to resort to mechanisms that can actually hold clouds with complex geometries and density distributions in approximate equilibrium (see also Burkert \& Hartmann 2004). Simulations of unbound clouds leading to inefficient star-formation were discussed  by e.g. Clark \emph{et al.} (2005 \& 2008). The Virial state of {\small GMCs} has also been interrogated numerically. Dobbs \emph{et al.} (2008) and Tasker \& Tan (2009) for instance, reported that clouds forming in a galactic disk typically have a Virial coefficient between 0.2 - 10. Furthermore, Dobbs \emph{et al.} (2011) argued that intercloud collisions in the disk of a galaxy was a viable mechanism to render clouds unbound.  \\ \\ 
Alternatively, several other authors examined the formation of clouds out of collisions between large-scale flows (e.g. V{\' a}zquez-Semadeni \emph{et al.} 1995; 2007, Joung \& Mac Low 2006). Hennebelle \emph{et al.} (2008), for instance, developed magnetohydrodynamic realisations of moderately supersonic atomic flows to demonstrate the formation of clouds. They demonstrated that the density {\small PDF}s of the resulting cloud were consistent with those reported observationally. Klessen \& Hennebelle (2010) argued that turbulence in the {\small ISM}   could be driven during the early phases of Galactic evolution when the disk was still probably accreting gas. They demonstrated this by setting-up a system of colliding-flows with typical accretional velocities of $\sim$15 km/s to $\sim$20 km/s. Their simulations showed that the collision induced a velocity dispersion $\sim$4 km/s - 5 km/s in the diffuse ($\lesssim$ 100 cm$^{-3}$) post-collision gas and a relatively weaker velocity dispersion, $\lesssim 1$ km/s, in dense gas ($\gtrsim 10^{4}$ cm$^{-3}$). \\ \\
In other similar work, Heitsch \emph{et al.} (2008 a) demonstrated that the post-collision cloud was never in any equilibrium though there was always an approximate equipartition of energy. The post-collision cloud, however, did develop localised centres of gravitational collapse. More recently, Carroll-Nellenback \emph{et al.} (2014) argued that collision between fractal flows was more likely to delay core-formation in the post-collision cloud. Likewise, Stanchev \emph{et al.} (2015) simulated a collision between uniform flows of atomic gas to reconcile the physical properties of the Perseus {\small MC}. The principal objective of their exercise was to investigate the length-scale over which gravity was likely to dominate turbulence. These efforts, however, neither address the issue about the observed variations in physical properties of clouds nor do they shed much light on the possibility that variations in the magnitude of interstellar pressure could possibly reconcile these observed variations in cloud-properties; see for instance Hughes \emph{et al.} (2013b).\\ \\ 
 Interaction between converging flows can be envisaged within the classic density-wave paradigm in which the arms of a typical spiral galaxy are believed to be patterns generated due to the propagation of a disturbance in the density field of a galactic-disk. In this paradigm molecular clouds are assembled in crests/troughs of the propagating wave where gas-flows converge (see for instance the recent review by Dobbs \& Baba 2014).  In this work we  therefore investigate the dynamical evolution of the post-collision cloud that is assembled via collision between such flows. We will investigate this problem numerically by developing self-gravitating realisations of flows having initially uniform density and colliding head-on; the case of non-headon collision between flows will be investigated in a sequel to this paper.
In particular, we will address the issue about the possible dependence of various physical properties and especially the size-linewidth relation, column density {\small N-PDF}, the gas surface density, $\Sigma_{gas}$, the magnitude of internal pressure, $P_{int}$, in a cloud and the time-scale on which gas in a cloud is assembled into the dense phase (the gas-depletion timescale), on the magnitude of the external pressure, $P_{ext}$, or equivalently, on the magnitude of the precollision velocity. The layout of the paper is as follows : In \S 2  we discuss the numerical method and the initial conditions for these simulations and present the results in \S 3. These results are discussed in \S 4 and we conclude in \S 5. 
%============================================================================

\begin{figure}
\label{Figure 1}
\vspace{1pc}
\centering
\includegraphics[angle=0,width=0.5\textwidth]{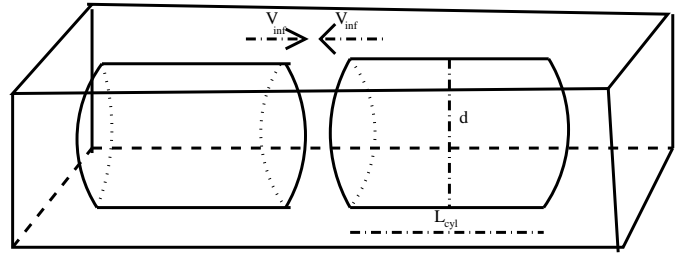}
\caption{Cartoon showing a schematic representation of a head-on collision between identical cylindrical gas-flows. Each flow has length, $L_{cyl}$, radius, $R_{cyl}=0.5*d$, and initial velocity magnitude, $V_{inf}$. See Table 1 for other physical details.}
\end{figure}
%----------------------------------------------------

\section{Numerical Method and Initial Conditions} 
\subsection{Initial conditions}
The set-up, as shown in Fig. 1, involves merely two identical cylindrical gas-flows of uniform density directed head-on towards each other and each having an equal magnitude of initial velocity, $V_{inf}$.  Individual flows are characterised by their mean density, $\bar{n}$, the pre-collision temperature, $T_{gas}$, while $M_{gas}$ is the total mass of gas in the computational box. Listed in Table 1 are details of the realisations developed in this work. The mass of the post-collision cloud, $M_{cld}$, which we nominally define as the volume of gas having density in excess of 50 cm$^{-3}$ will be $\lesssim M_{gas}$. We performed 11 realisations of the problem to produce a magnitude of pressure, $P_{ext}$, confining the post-collision cloud in the range 10$^{3}$ K cm$^{-3}$ - 10$^{8}$ K cm$^{-3}$ to mimic the ambient conditions that prevail at different radial locations in the Galactic disk. The least massive clouds and the lowest magnitude of the interstellar pressure, $P_{ext}$ (typically, 10$^{4}$ K cm$^{-3}$), is found in the outermost regions of the Galactic disk. On the contrary, the most massive clouds appear to be preferably located in the inner regions of the Galactic disk where the magnitude of $P_{ext}$ is comparatively higher ($\gtrsim 10^{5}$ K cm$^{-3}$) (Rice \emph{et al.} 2016; Kasparova \& Zasov 2008). The most massive clouds ($\gtrsim 10^{5}$ M$_{\odot}$), confined by a relatively large magnitude of interstellar pressure ($P_{ext}\gtrsim 10^{6}$ K cm$^{-3}$), are found in the Galactic Central molecular Zone ({\small CMZ})(e.g. Ao \emph{et al.} 2013, Rathborne \emph{et al.} 2014). Listed in column 4 of  Table 1 is the magnitude of $P_{ext}$ in each realisation of this numerical exercise.
%-----------------------------------------------------------------------------
\begin{table*}
 \centering
 \begin{minipage}{150mm}
  \caption{Physical details of realisations.}
  \begin{tabular}{|l|l|l|l|l|l|l|}
\hline
Serial & Physical &  Pre-collision Velocity & $\Big(\frac{P_{ext}}{k_{B}}\Big)$ & M$_{gas}$ & M$_{min}$\footnote{As defined by Bate \& Burkert (1997)} & Comment \\
No. &  Properties & of gas-flows, V$_{inf}$ [(km/s)] & [K cm$^{-3}$] & [M$_{\odot}$] &  [M$_{\odot}$] & \\
&& Pre-collision Mach  &&&& \\
&& number ($\mathcal{M}$) &&&& \\
\hline
1 & $L_{cyl}$ = 130 pc, $R_{cyl}$ = 14 pc & 4.8 (0.7) & 2.8$\times 10^{3}$ & 4.3$\times 10^3$ & 0.2 & Warm Atomic \\
& $\bar{n}$ = 1 cm$^{-3}$, $T_{gas}$ = 5000 K & & & & & Flow($\mu$ =1 amu)  \\
& $\bar{n}_{max}\sim 10^{6}$ cm$^{-3}$, $h_{avg}^{init}\sim$ 0.2 pc &&&&& \\
\hline
2 & Same as (1) & 9.0 (1.4) & 9.8$\times 10^{3}$ & 4.3$\times 10^3$ & 0.2 & Warm Atomic \\
& & & & & & Flow \\
\hline
3 & Same as (1) & 19.28 (3.0) & 4.45$\times 10^{4}$ & 4.3$\times 10^3$ & 0.2 & Warm Atomic \\
& & & & & & Flow \\
\hline
4 & $L_{cyl}$ = 50 pc, $R_{cyl}$ = 10 pc & 3.45 (1.7) & 7.3$\times10^{4}$ & 4.2$\times 10^4$ & 2.0 & Warm Atomic \\
& $\bar{n}$ = 50 cm$^{-3}$, $T_{gas}$ = 500 K & & & & & Flow($\mu$ =1 amu)  \\
& $\bar{n}_{max}\sim 10^{5}$ cm$^{-3}$, $h_{avg}^{init}\sim$ 0.13 pc &&&&& \\
\hline
5 & Same as (4) & 6.1 (3.0)& 2.25$\times10^{5}$ & 4.2$\times 10^4$ & 2.0 & Warm Atomic \\
& & & & & &  Flow  \\
\hline
6 & Same as (4) & 10.36 (5.1) & 6.51$\times10^{5}$ & 4.2$\times 10^4$ & 2.0 & Warm Atomic \\
& & & & & &  Flow  \\
\hline
7 & $L_{cyl}$ = 30 pc, $R_{cyl}$ = 10 pc & 5.6 (15.1) & 9.0$\times10^{5}$ & 1.2$\times 10^5$ & 6.0 & Cold Molecular \\
& $\bar{n}$ = 100 cm$^{-3}$, $T_{gas}$ = 40 K & & & & & Flow($\mu$ =2.29 amu)  \\
& $\bar{n}_{max}\sim 10^{4}$ cm$^{-3}$, $h_{avg}^{init}\sim$ 0.1 pc &&&&& \\
\hline
8 & Same as (7) & 7.43 (20.0)& 1.6$\times10^{6}$ & 1.2$\times 10^5$ & 6.0 & Cold Molecular \\
& & & & & &  Flow  \\
\hline
9 & Same as (7) & 14.86 (40.0)& 6.4$\times10^{6}$ & 1.2$\times 10^5$ & 6.0 & Cold Molecular \\
& & & & & &  Flow  \\
\hline
10 & Same as (7) & 29.7 (79.97)& 2.56$\times10^{7}$ & 1.2$\times 10^5$ & 6.0 & Cold Molecular \\
& & & & & &  Flow  \\
\hline
11 & Same as (7) & 55.6 (149.7)& 8.9$\times10^{7}$ & 1.2$\times 10^5$ & 6.0 & Cold Molecular \\
& & & & & &  Flow  \\
\hline
\end{tabular} \\
\end{minipage}
\end{table*}
%--------------------------------------------------------------------------

\begin{figure*}
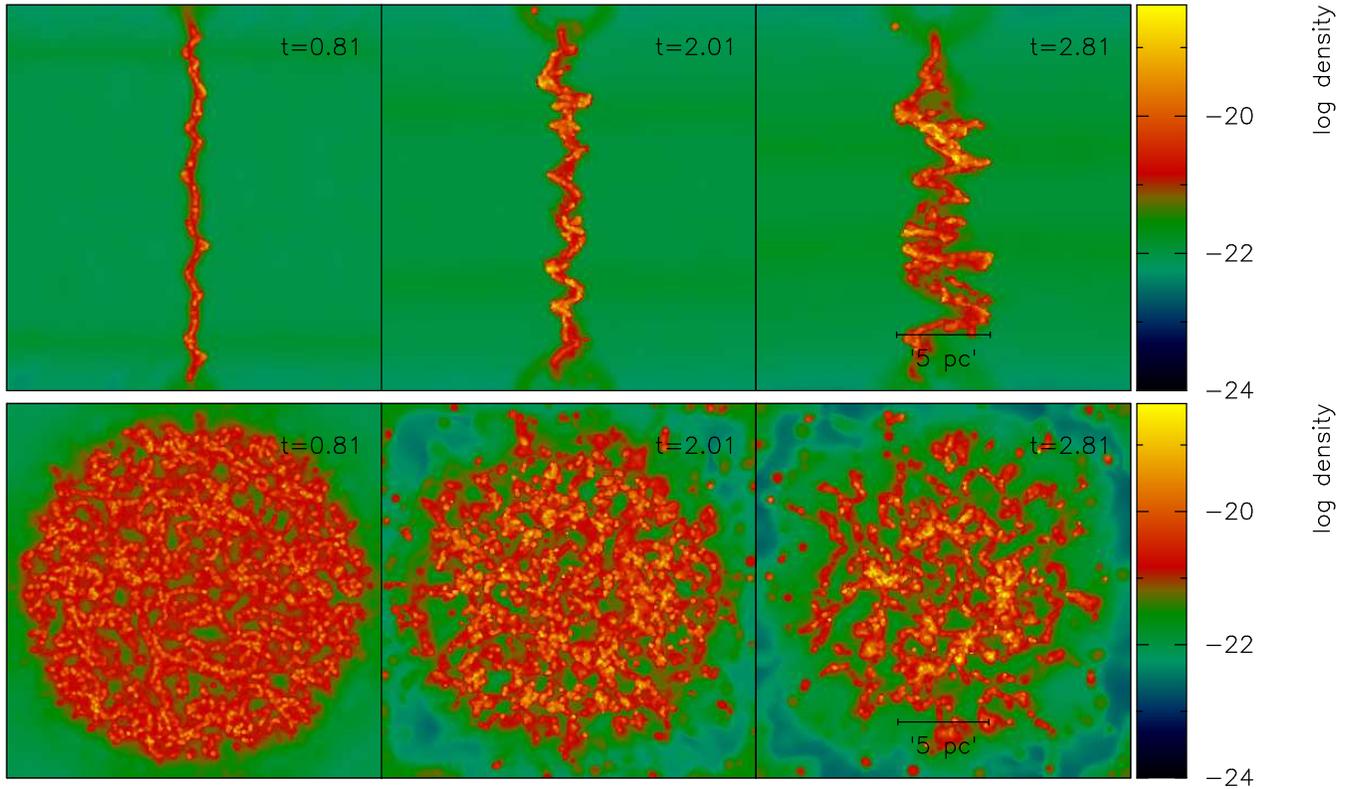

\label{Figure 2}
\vspace{1pc}
\centering
\includegraphics[angle=270,width=\textwidth]{N50P3XYrendens.eps}
\includegraphics[angle=270,width=\textwidth]{N50P3XZrendens.eps} 
\caption{Shown here are the rendered density images of the shocked-slab in realisation 6 at different epochs; time in Myrs has been marked on the top right-hand corner of each panel. Pictures on the upper-panel show the post-collision slab in the plane of collision while those on the lower-panel show its transverse section taken through the mid-plane. That the growth of the shell-instability causes the slab to buckle and fragments it rapidly is evident from these images.}
\end{figure*}

%-----------------------------------------------------------------------------
\subsection{Numerical Method}
Realisations discussed in this work were developed using the well tested {\small SPH} code {\small SEREN} (Hubber \emph{et al.} 2011). We used the Monaghan-Riemann viscosity (Monaghan 1997), to capture shocks in the simulations discussed below. Signal velocity in the viscous dissipation term was calculated as
\begin{equation}
v_{sig}(i,j) = (c_{i}^{2} + \beta(\textbf{v}_{ij}\cdot\hat{\textbf{r}}_{ij})^{2})^{1/2} + (c_{j}^{2} + \beta(\textbf{v}_{ij}\cdot\hat{\textbf{r}}_{ij})^{2})^{1/2} - \textbf{v}_{ij}\cdot\hat{\textbf{r}}_{ij},
\end{equation} 
 where $(c_{i},c_{j})$ are the respective sound-speeds for particles $(i,j)$, $\textbf{v}_{ij}\equiv\vert \textbf{v}_{i}- \textbf{v}_{j}\vert$ and $\hat{\textbf{r}}_{ij}$ is the unit vector along the direction $\textbf{r}_{ij}$, connecting the particles $(i,j)$. This expression for $v_{sig}(i,j)$, though similar to the conventional prescription, performs better for stronger shocks (Monaghan 1997; Toro 1992). In other words, the resulting shocks are sharper. Dynamical cooling of gas, as in some of our earlier work, was implemented with the aid of a parametric cooling-curve for the interstellar medium (e.g. Koyama \& Inutsuka 2002; Vazquez-Semadeni \emph{et al.} 2007). \\
\textbf{\emph{Resolution}} \\
The number of gas particles, $N_{gas}$, used in a realisation were calculated such that the Bate-Burkert criterion for resolution was satisfied at the minimum gas temperature ($\sim$10 K), set for each realisation. The minimum resolvable mass in an {\small SPH} realisation is
\begin{equation}
M_{min}\sim\Big(\frac{2N_{neibs}}{N_{gas}}\Big)\cdot M_{gas}
\end{equation}
Bate \& Burkert (1997); here $N_{neibs}$=50, is the fixed number of neighbours that each {\small SPH} particle has and, $M_{gas}$, the mass of gas in the computational domain. We note that the minimum resolvable mass, $M_{min}$, in this exercise has been varied between 0.2 M$_{\odot}$ and 6 M$_{\odot}$ for the three choices of initial gas density. This was done to merely keep the number of gas particles, $N_{gas}$, within manageable limits. In this work we used $N_{gas}\sim 2.15\times 10^{6}$ and $N_{icm}=0.85\times 10^{6}$ (those representing the intercloud medium, {\small ICM}, which like the live gas-particles exert hydrodynamic force, but unlike them do not posses self-gravity), particles to develop each realisation. Consequently, the initial average smoothing length, $h_{avg}^{init}$, that determines the extent of the smallest resolvable spatial-region, also varies. It is defined as
\begin{displaymath}
h_{avg}^{init} = \Big(\frac{N_{neibs}}{N_{gas}}\Big)\Big(\frac{3d^{2}L_{cyl}}{128}\Big),
\end{displaymath}
$d$, and, $L_{cyl}$, being respectively the diameter and the length of individual flow; see cartoon in Fig. 1. \\ \\
The {\small ICM} particles were assembled to jacket the cylindrical flows and were set-up such that there was no density contrast across the gas-{\small ICM} interface. The entire assembly was then placed in a periodic-box meant only to ghost particles in the box, i.e. particles leaving from one face were allowed to re-enter from the opposite face. 
Finally, listed in  column 2 of Table 1 are the physical details of the pre-collision flows in each realisation. We reiterate, our extant interest lies in investigating the physical properties of the assembled cloud and the variation of their magnitude as a function of the magnitude of external pressure, $P_{ext}$, and on the efficiency with which gas in these clouds is cycled into potentially star-forming pockets. Consequently, we do not follow the actual formation of prestellar cores in these realisations and defer this aspect of the question to a sequel to this article. In spite of the relatively coarse numerical resolution, these realisations are sufficiently well resolved as to render the deduced properties of assembled clouds robust.
%------------------------------------------------------------------

%-------------------------------------------------------------
\section[]{Results}
\subsection{Evolution of the shocked-slab}
Precollision flows were supersonic in all the simulations listed in Table 1, care for the first realisation where the individual gas-flows were sub-sonic with a precollision Mach number of 0.7. In an earlier work (Anathpindika 2009a,b), we have demonstrated the difference in evolution of a post-collision slab confined by shocks generated by a 
a super-sonic collision, as against one confined by ram-pressure, as in the case of a sub-sonic collision. The slab in the former case is thin and soon after its assembly, develops corrugations on its surface which marks the onset of the Thin Shell Instability ({\small TSI}). The amplitude of the associated crests and troughs grows rapidly due to the transfer of momentum between them and causes the shocked-slab to buckle. \\ \\
Soon after it is triggered, the {\small TSI} grows non-linearly, a phase better known as the the Non-linear Thin Shell Instability ({\small NTSI}), and rapidly fragments the slab. Shortly thereafter, the amplitude of slab-oscillations becomes comparable to the thickness of the slab as the {\small NTSI} saturates (Vishniac 1994, Heitsch \emph{et al.} 2008a, Anathpindika 2009a). Importantly, growth of the {\small NTSI} triggers a strong shearing motion between slab-layers which leads to mixing between them and dissipates turbulent energy that is crucial towards supporting the slab against self-gravity. The upshot being, the shocked-slab soon puffs-up as the growth of the {\small NTSI} approaches saturation and the slab eventually loses support against self-gravity. Consequently, the slab collapses globally to form a dense elongated globule in the plane of collision.\\ \\
By contrast, the ram-pressure confined post-collision slab evolves in quasi-static fashion, via the interplay between self-gravity and the kinetic energy injected by the collision (Anathpindika 2009b). Besides, in either case, the slab is also attended to by the cooling instability. Heitsch \emph{et al.} (2008b) for instance, discussed the parameter regimes over which the respective instabilities are likely to dominate. Within the context of this work it is important to underline the ability of the {\small SPH} to model the {\small TSI} and its non-linear excursion, the {\small NTSI}. Anathpindika (2009 a) and Hubber \emph{et al.} (2013) demonstrated that the {\small SPH} can indeed reproduce the analytically predicted growth-rate by Vishniac (1994) for this instability. \\ \\
By choice, all but one realisation in this work involve initially supersonic gas-flows suggesting that the post-collision slab in these slabs would be attended by the {\small NTSI}, albeit the triggering of this instability is delayed for relatively lower pre-collision Mach numbers. Here we therefore present images of the shocked slab from only one realisation. Shown in Fig. 2  is the observed temporal evolution of the shocked slab in realisation 6. Buckling along the slab-surface and the associated temporal amplification of the corrugations on its surface, both classic identifying features of the {\small NTSI}, are visible in the rendered plots shown on the upper-panel of Fig. 2. Subsequently, as  it is evident from the picture in the right-hand frame ($t\sim$2.81 Myrs), on the upper-panel of Fig. 2, the amplitude of perturbations on slab-surface becomes comparable to its thickness when growth of the {\small NTSI} saturates. Similarly, pictures on the lower-panel of this figure show the fragmentation in the plane of the shocked-slab. With passage of time the slab becomes dominated by a network of dense filaments.  However, we did not follow the calculations in this realisation to the stage where the shocked-slab collapses, though we did so for a latter case, listed 10 in Table 1 simply because a shocked-slab such as the one in this latter case is known from our earlier works (Anathpindika 2009 a), to evolve on a relatively shorter timescale.
%--------------------------------------------------------

\begin{figure}
\label{Figure 3}
\vspace{0pc}
\centering
\includegraphics[angle=270,width=0.5\textwidth]{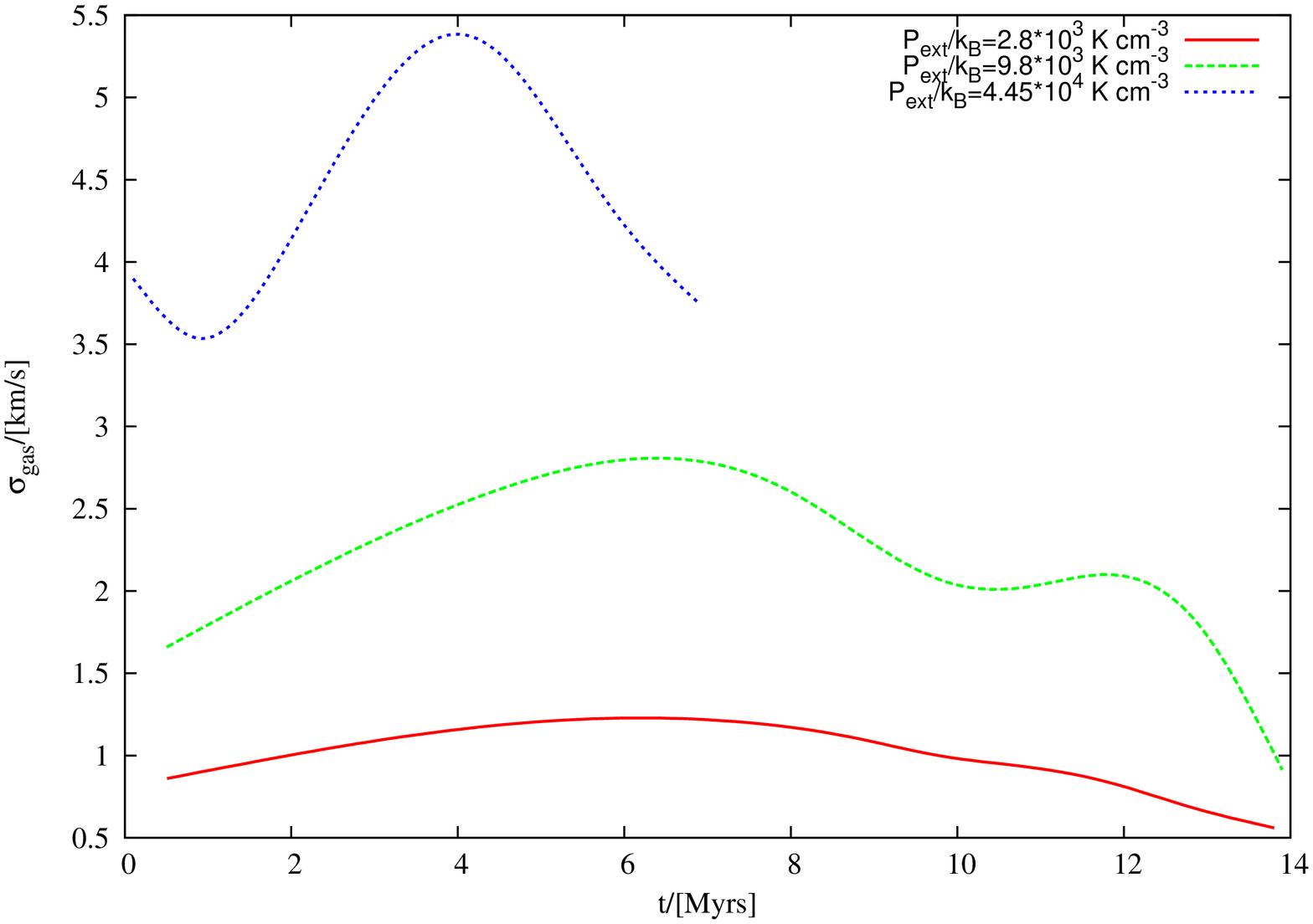}
\includegraphics[angle=270,width=0.5\textwidth]{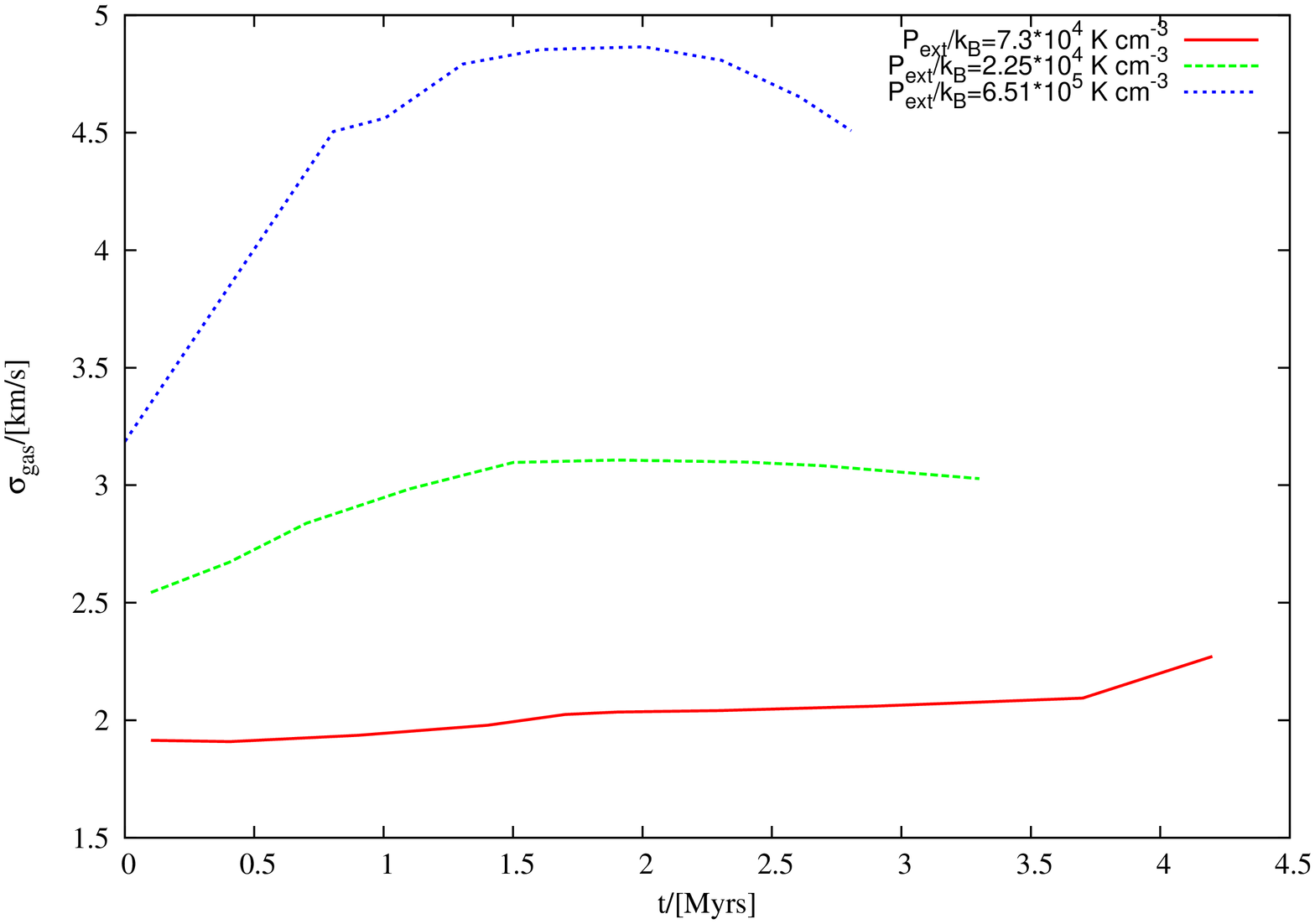}
\includegraphics[angle=270,width=0.5\textwidth]{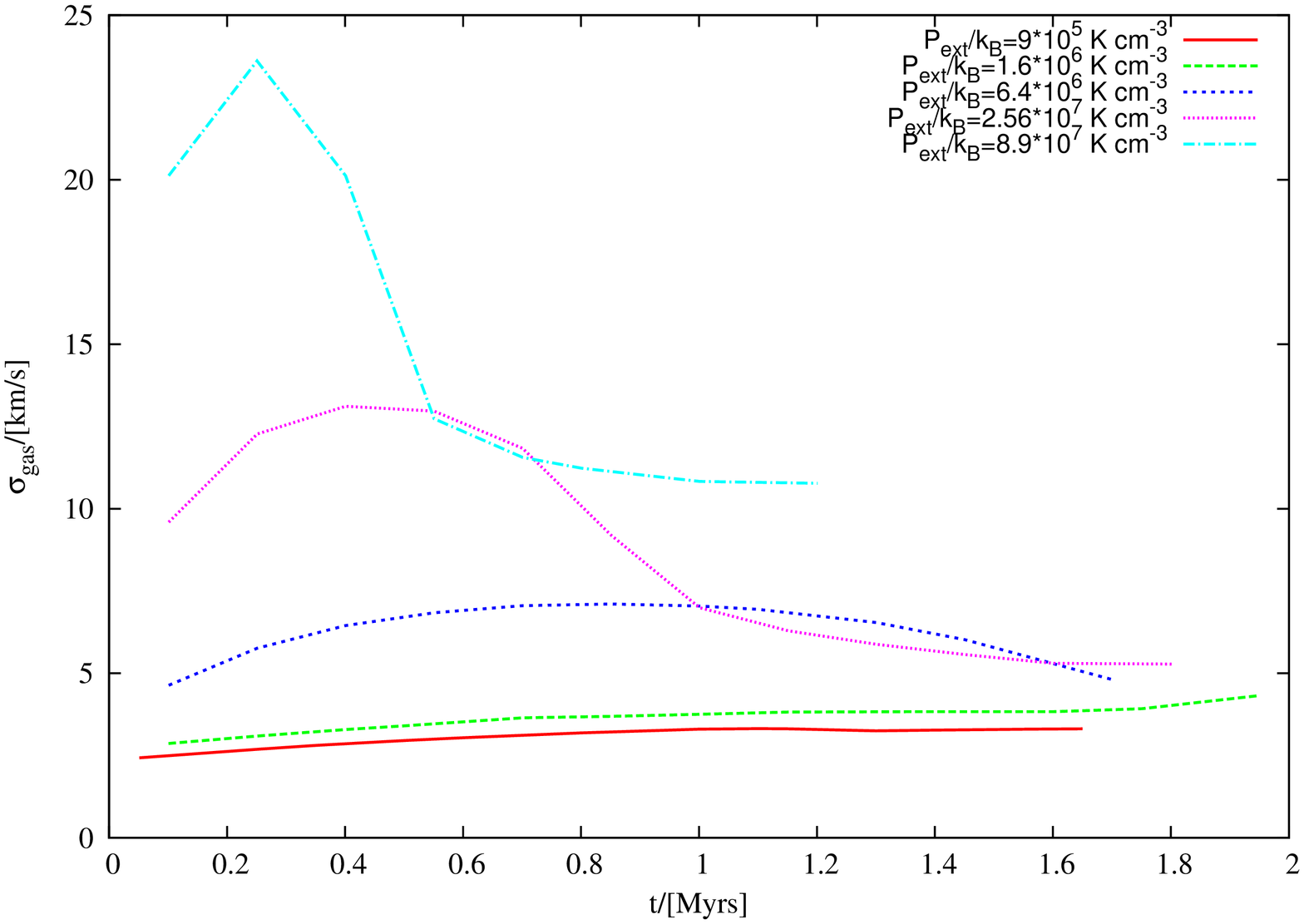}
\caption{Velocity-dispersion for gas in the shocked-slab for each set of simulations with initial choice of gas density, $\bar{\mathrm{n}}$=1 cm$^{-3}$, 50 cm$^{-3}$ and 100 cm$^{-3}$ is shown respectively on the upper, middle and the lower-hand panel.}
\end{figure}
%-------------------------------------------------------------
\subsection{Typical diagnostics of gas in a post-collision slab}
\textbf{Gas-velocity dispersion} \\
Growth of dynamic instabilities injects a velocity field in the slab-layers and shown in various panels of Fig. 3 is the temporal variation of velocity dispersion, $\sigma_{gas}$, of gas in the post-collision slab for the three choices of initial gas-density viz., $\bar{\mathrm{n}} = $1 cm$^{-3}$ (realisations 1-3), 50 cm$^{-3}$ (realisations 4-6) and 100 cm$^{-3}$(realisations 7-11). Irrespective of the choice of the initial gas-density, the plots exhibit two common features : (\textbf{i}) the velocity dispersion acquires a maxima as the post-collision slab steadily accretes gas soon after its formation. Thereafter, as the shell-instability begins to grow on the slab-surface, it causes kinetic energy within the slab-layers to dissipate as manifested by a gradual tapering-off of the velocity-dispersion at later times of evolution of the slab, and (\textbf{ii}) despite this dissipation, the magnitude of velocity-dispersion, $\sigma_{gas}$, in general is proportional to the magnitude of the precollision velocity or equivalently, to the magnitude of the external pressure, $P_{ext}$; see column 4 of Table 1. Here we also draw the attention of our reader to the fact that the realisations 1-3, as can be seen from the plot in the upper-panel of this figure, were allowed to run for significantly longer than other cases simply to allow the gas time to cycle into the dense phase i.e., to become potentially star-forming. Consequently, we also see the steep decline in magnitude of the gas velocity dispersion, $\sigma_{gas}$, at later epochs in these plots. Similar steep decline is also visible in cases where the respective realisations were allowed to run to the point where the post-collision slab had begun collapsing. At earlier epochs though, the respective curves in all the sets look mutually similar. \\ \\
%----------------------------------------------------------------
\begin{figure*}
\label{Figure 4}
\vspace{1pc}
\centering
\includegraphics[angle=270,width=0.45\textwidth]{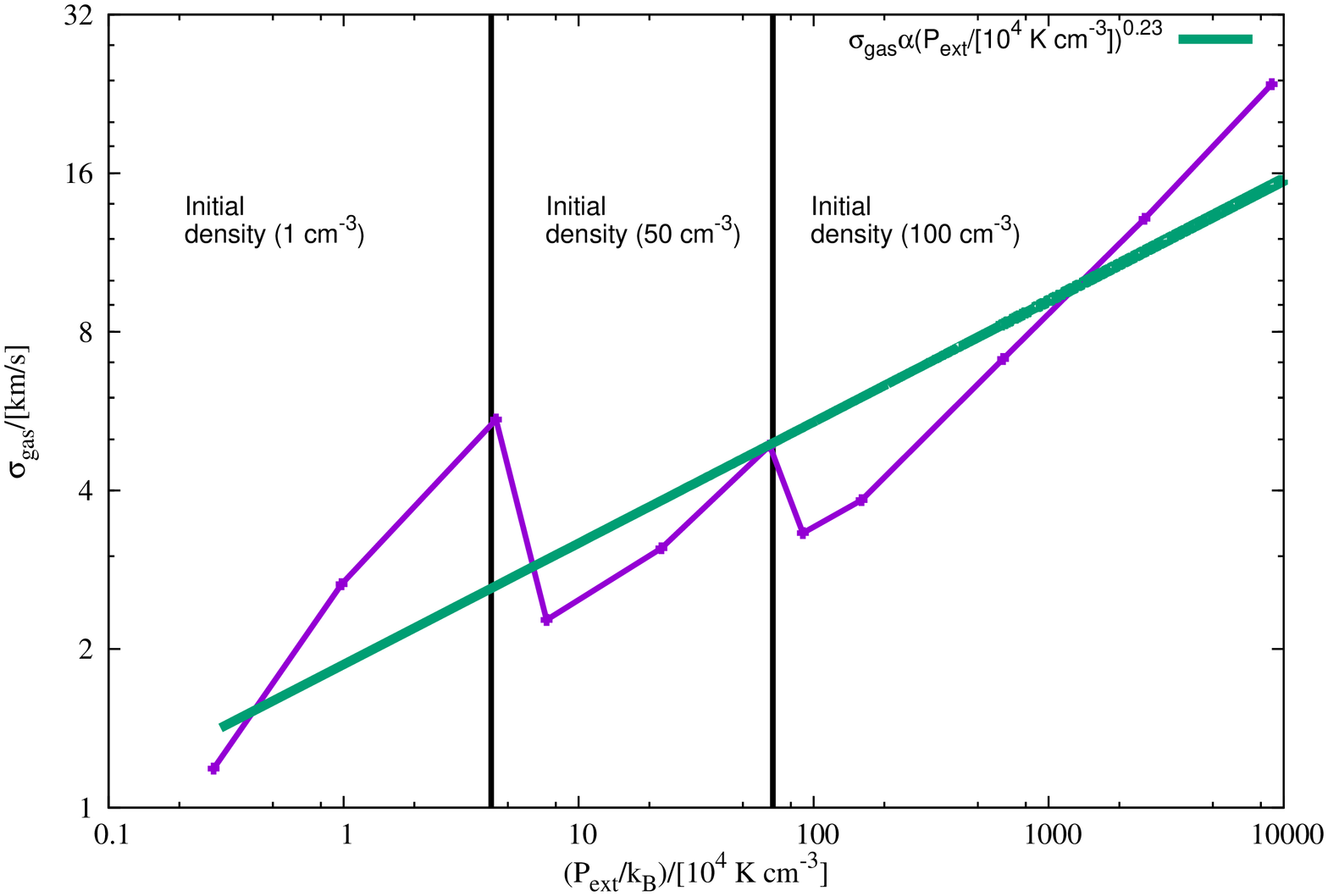}
\includegraphics[angle=270,width=0.45\textwidth]{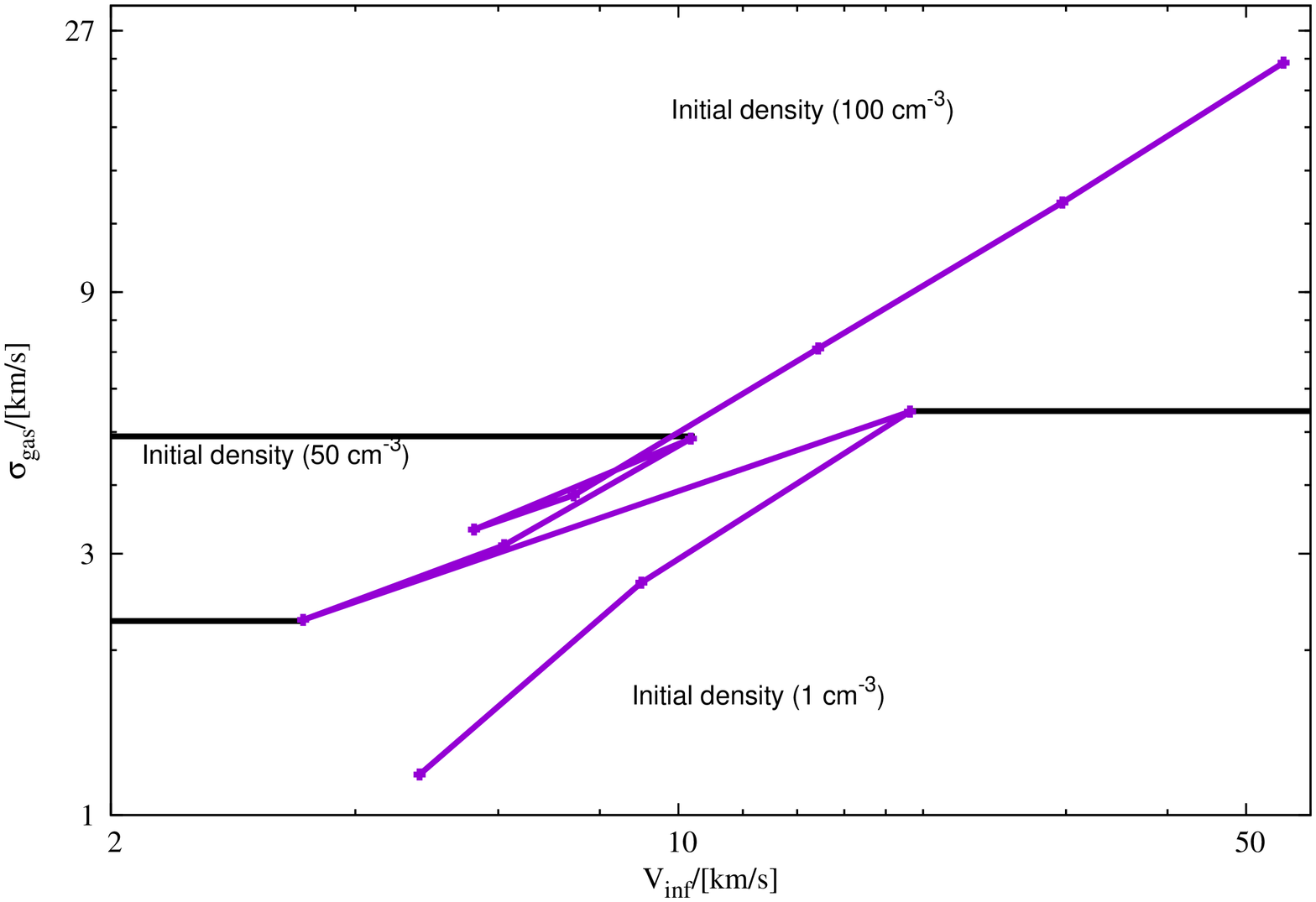}
\caption{Shown respectively on the left and the right-hand panels here are plots showing the maximum magnitude of velocity dispersion, $\sigma_{gas}$, for each realisation against the respective magnitude of external pressure, $P_{ext}$, and, $\sigma_{gas}$, against the magnitude of the inflow velocity,V$_{inf}$, of gas.}
\end{figure*}
%----------------------------------------------------------------
Furthermore, shown on the left-hand panel of Fig. 4 is a plot of the magnitude of maximum velocity dispersion measured for each realisation i.e., the peak of each characteristic shown on the upper, middle and the lower-panel of Fig. 3, against the respective magnitude of external pressure, $P_{ext}$. Here one can readily see that a larger magnitude of external pressure, $P_{ext}$, induces a higher magnitude of velocity dispersion within the slab layers. The magnitude of velocity dispersion increases steadily for a given choice of the initial gas-density. In general, the data over the entire range of pressure can be fitted reasonably well with a power-law of the kind $\sigma_{gas}\propto (P_{ext}/[10^{4} \mathrm{K\ cm}^{-3}])^{0.23}$, derived by the technique of regression and which the reader will easily recollect, is roughly consistent with the power-law suggested by Elmegreen (1989); see Eqn. (1) above. This similarity in the exponent lends credence to the hypothesis that clouds merely represent energy-equipartition (e.g. Elmegreen 1989, Ballesteros-Paredes 2006, Heitsch \emph{et al.} 2008 a).  \\ \\
Similarly, shown on the right-hand panel of this figure is a plot showing $\sigma_{gas}$ for different choices of the inflow velocity, V$_{inf}$, across the realisations developed in this work. This plot is interesting because not only does it show that the magnitude of velocity dispersion, $\sigma_{gas}$, increases approximately linearly with increasing $V_{inf}$, but also helps reconcile the discontinuities in the magnitude of velocity-dispersion across density regimes in the $\sigma_{gas}-P_{ext}$  plot. The $\sigma_{gas} - V_{inf}$ plot reinforces the conclusion from our earlier work (Anathpindika 2009 a,b), that the magnitude of $\sigma_{gas}$ is fundamentally sensitive to the magnitude of V$_{inf}$. Pressure, on the other hand being a derived physical quantity, also depends on the mean density of the in-flowing gas so that the magnitude of pressure experienced by a cloud need not be the result of an interaction between gas-flows having a unique mean density and inflow velocity. Thus, in view of the plot on the left-hand panel of Fig. 4 we suggest, clouds experiencing relatively low to intermediate magnitudes of external pressure, $P_{ext} \sim 10^{3.5}$ K cm$^{-3}$ - 10$^{5.5}$ K cm$^{-3}$, may not show an unambiguous trend of increasing magnitude of gas-velocity dispersion, $\sigma_{gas}$, corresponding to an increasing magnitude of external pressure. Albeit, the trend is clearer at significantly higher magnitudes of $P_{ext}$, typically in excess of 10$^{6}$ K cm$^{-3}$, usually found in clouds closer to the Galactic centre (e.g. Rice \emph{et. al.} 2016). \\ \\
%-----------------------------------------------------------------------
\begin{figure*}
\label{Figure 5}
\vspace{1pc}
\centering
\includegraphics[angle=270,width=0.45\textwidth]{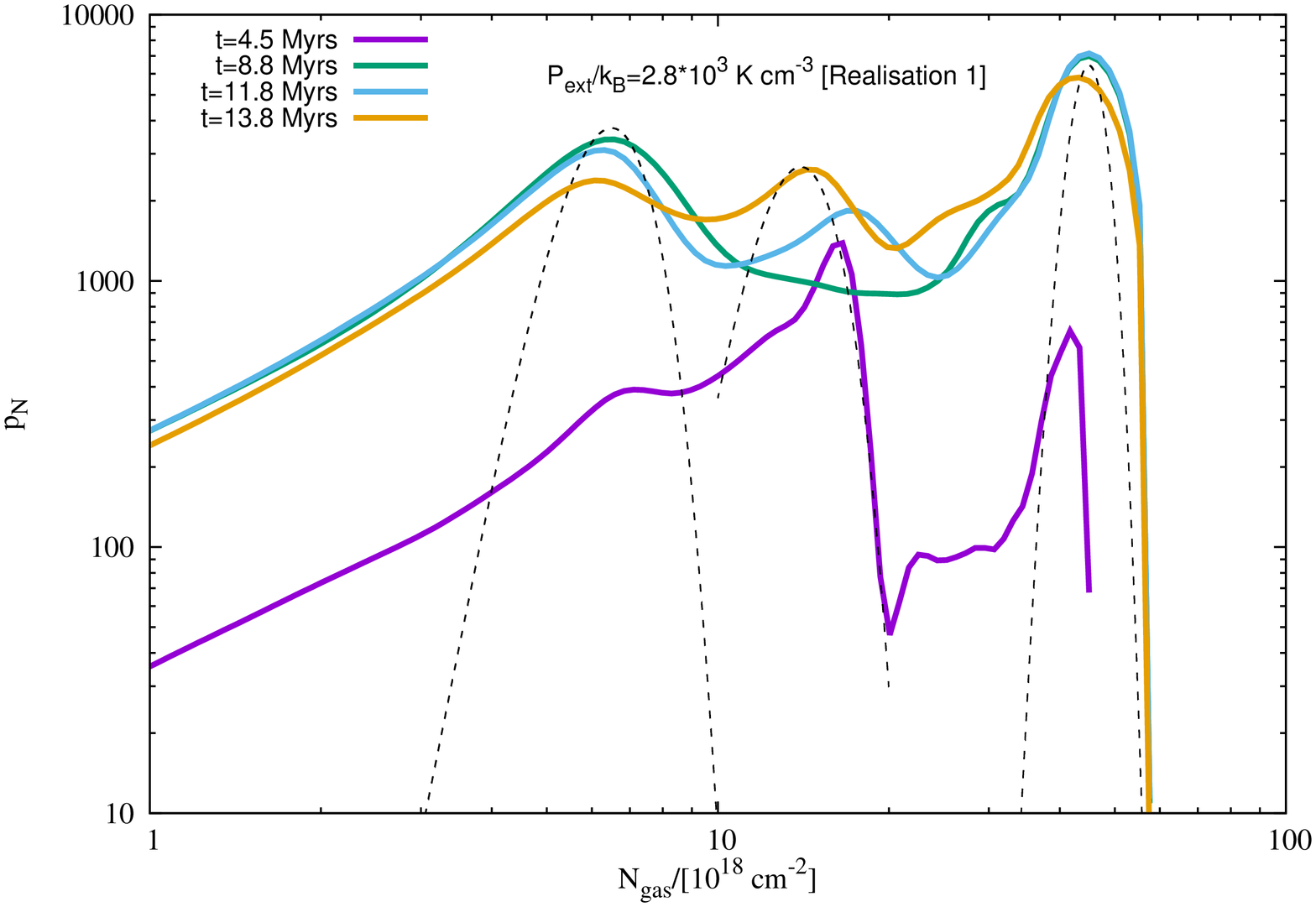}
\includegraphics[angle=270,width=0.45\textwidth]{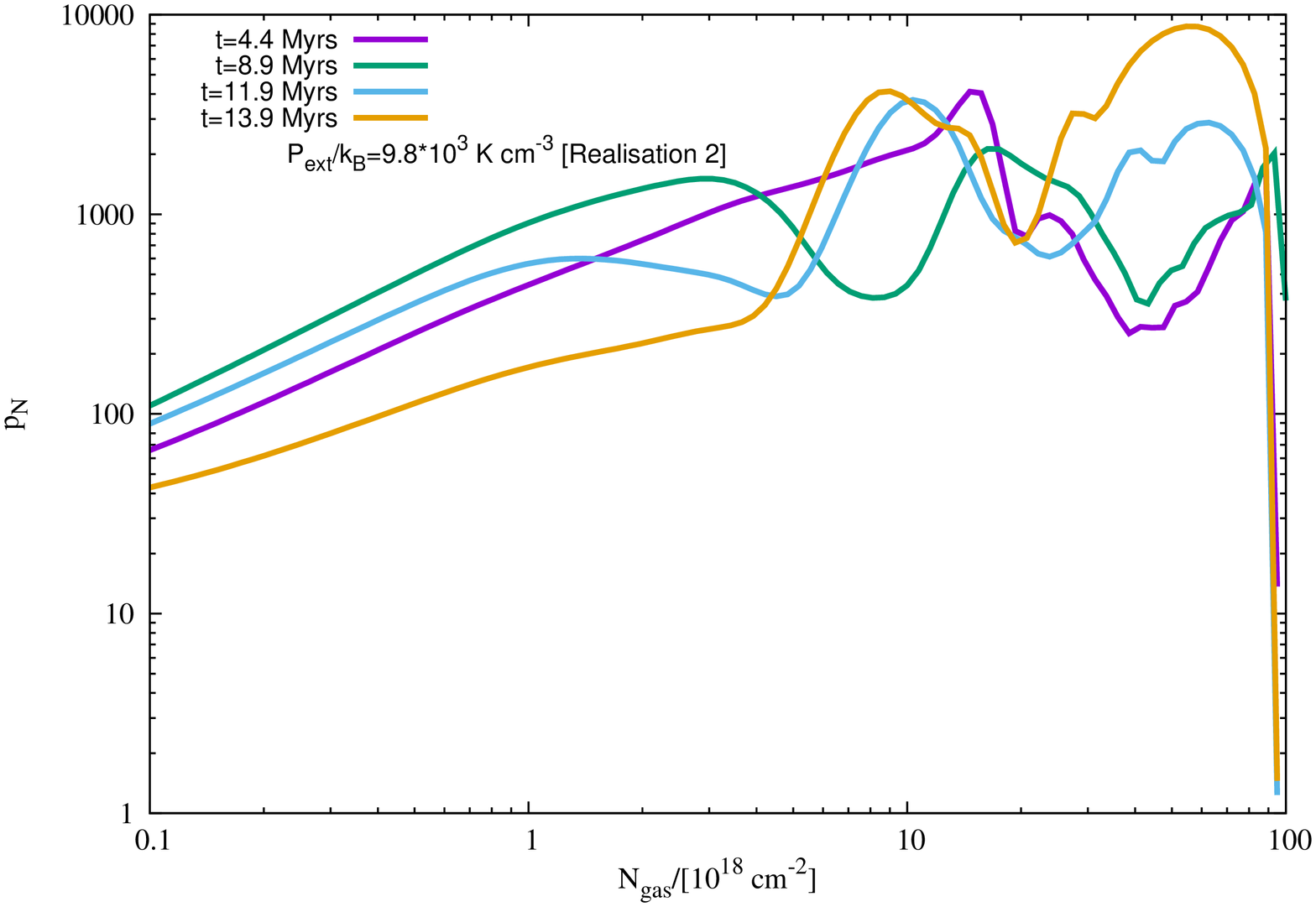}
\includegraphics[angle=270,width=0.45\textwidth]{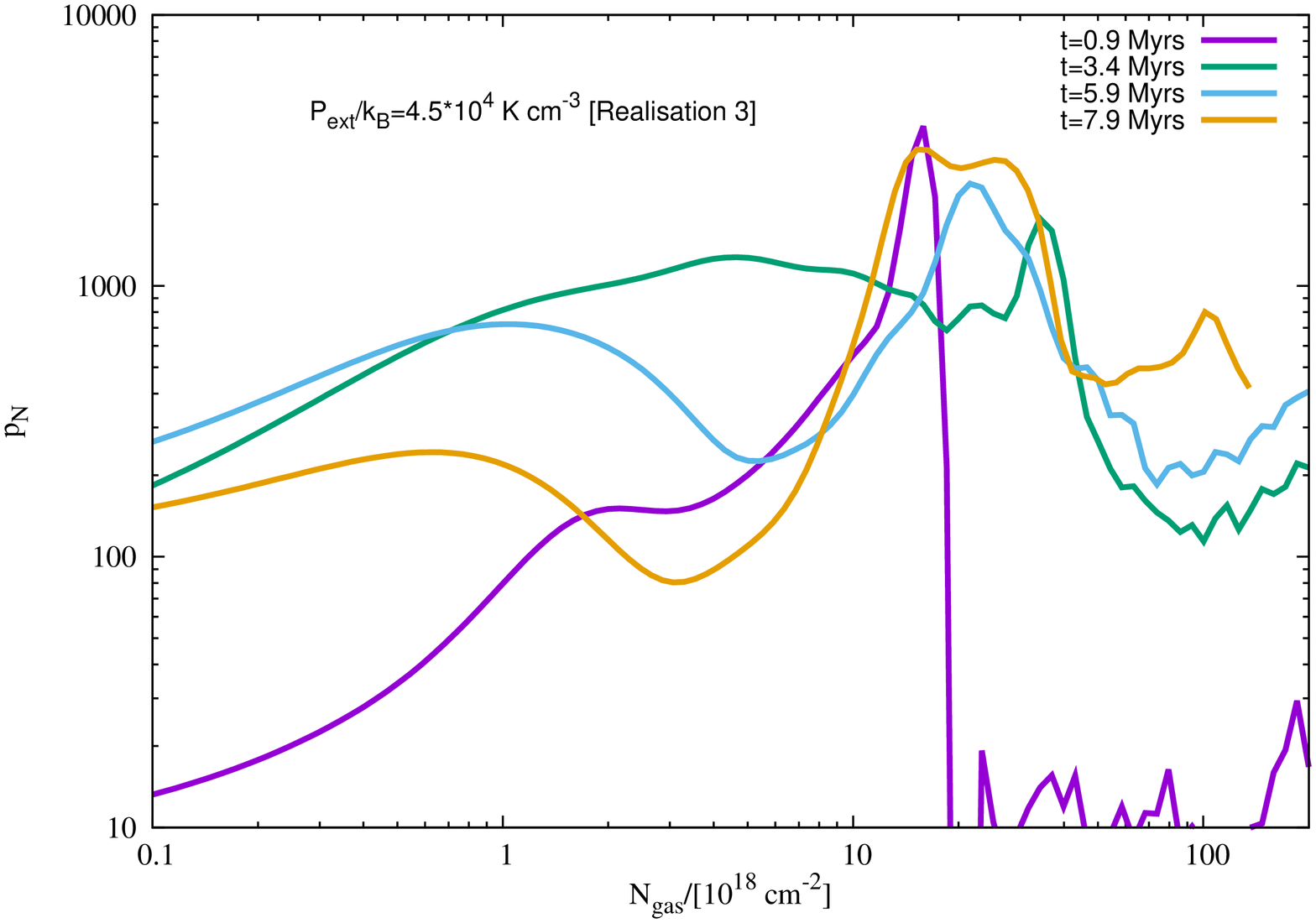}
\includegraphics[angle=270,width=0.45\textwidth]{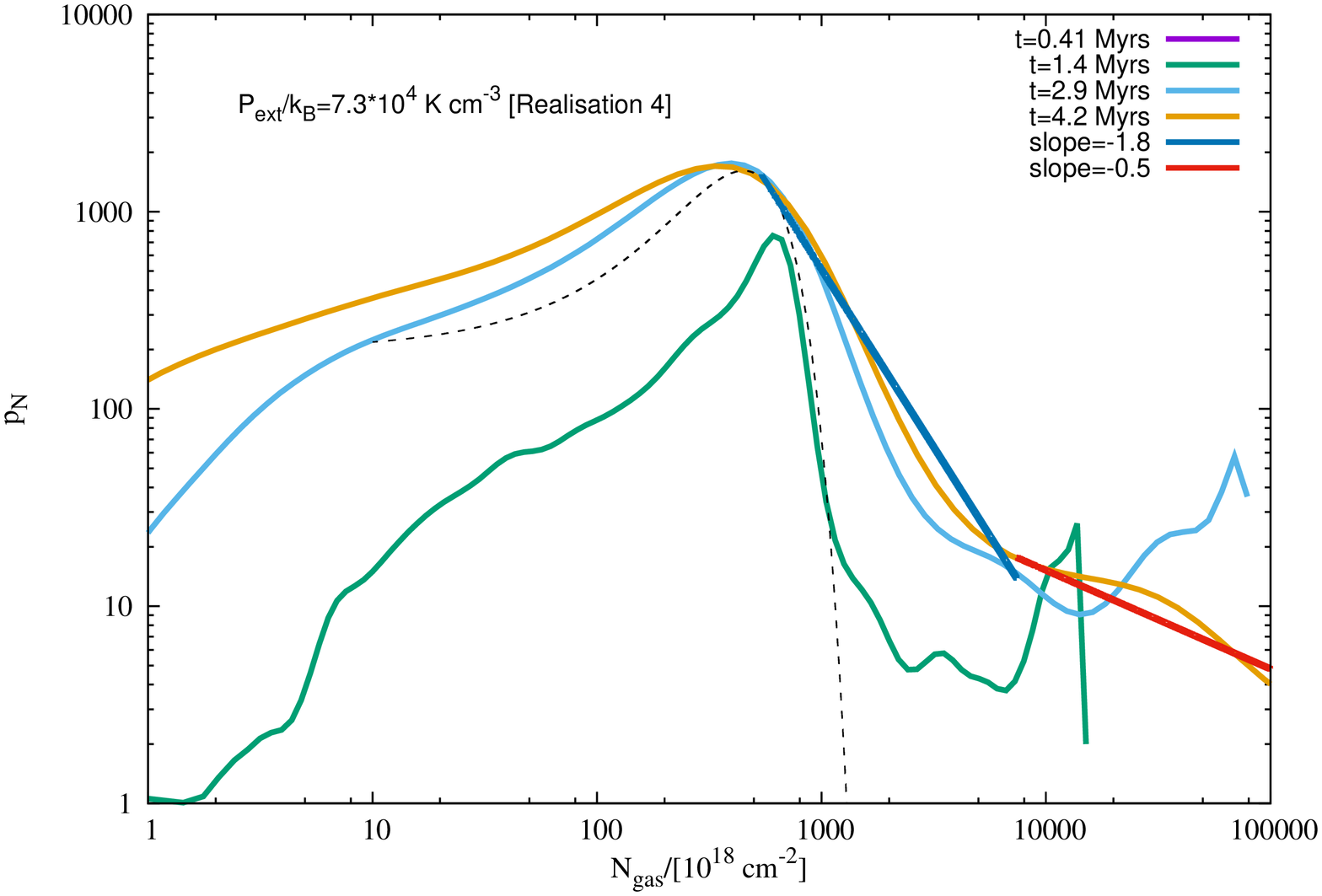}
\includegraphics[angle=270,width=0.45\textwidth]{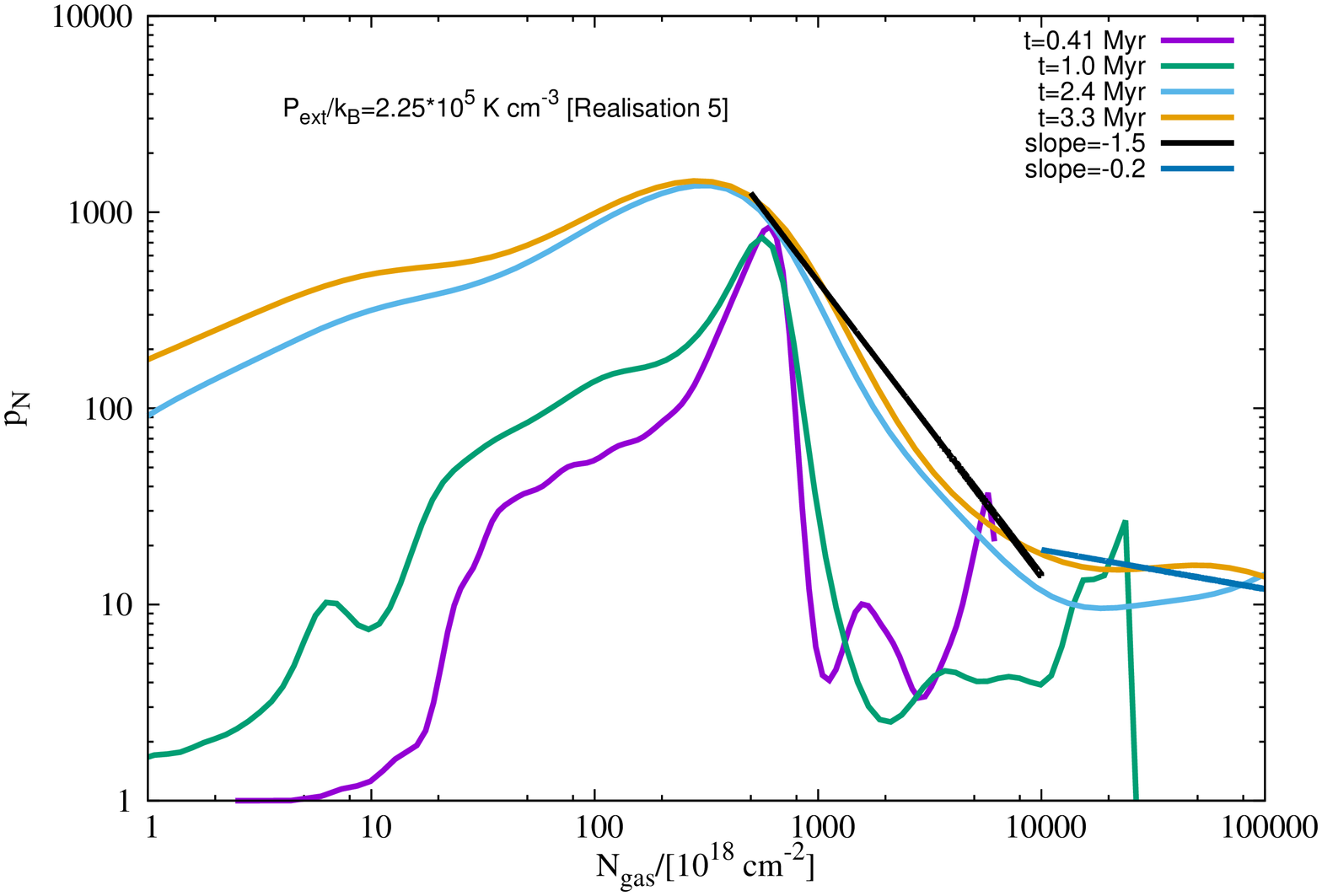}
\includegraphics[angle=270,width=0.45\textwidth]{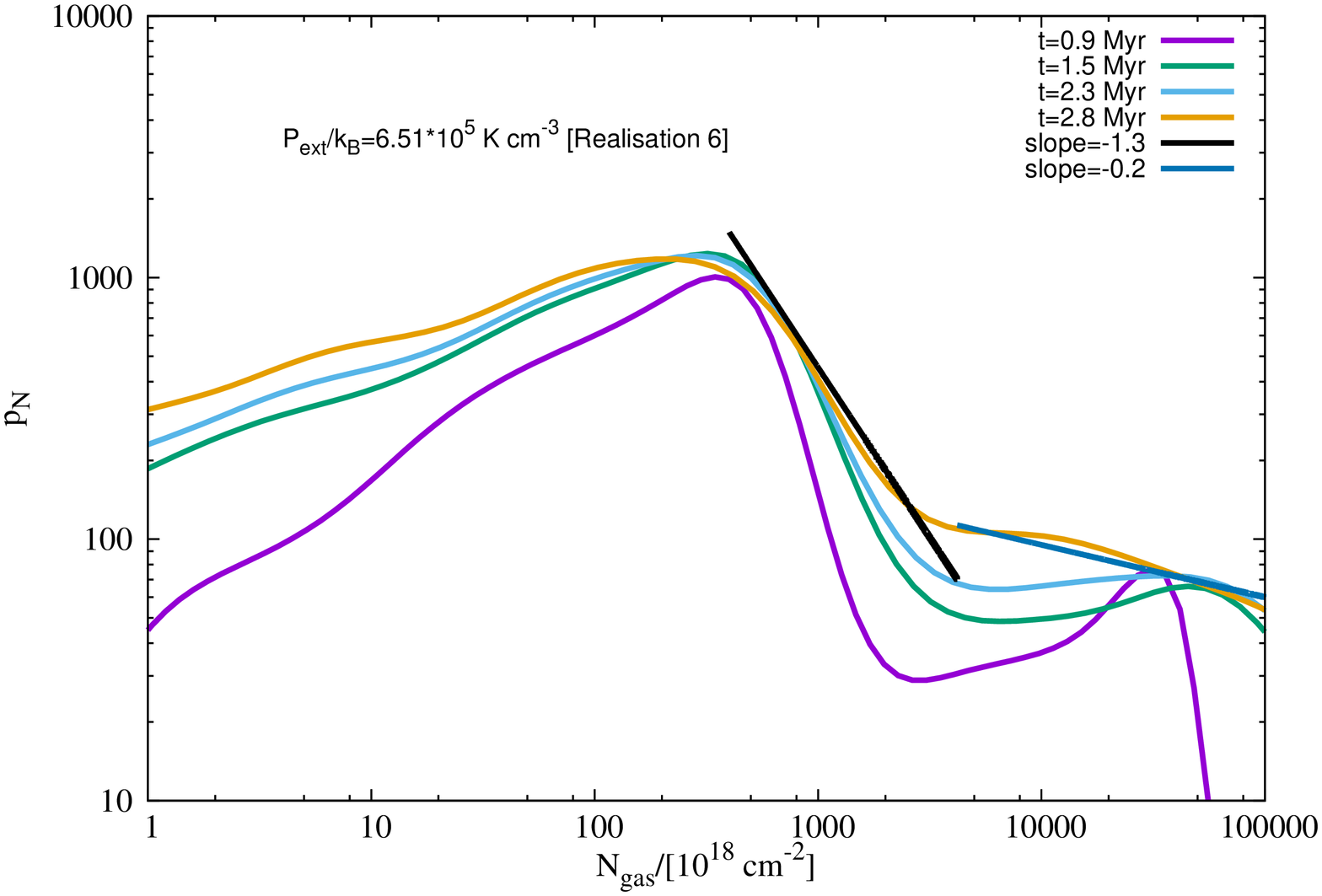}
\caption{Column density distributions ({\small N-PDF}s) at different epochs of the post-collision slab in realisations 1-3 (\emph{upper-panel} and \emph{central-panel} left-hand), and  realisations 4-6 (\emph{central-panel} right-hand and \emph{lower-panel}). The constituting lognormal distributions of a {\small N-PDF} have been shown on some plots using a black dashed line. The development of a power-law tail at higher densities, especially at latter epochs, can be readily seen in the {\small N-PDF}s for the latter realisations where the magnitude of external pressure was an order of magnitude higher in comparison to the former set of realisations.}
\end{figure*}
%
%------------------------------------------------------------------------
\textbf{Column density probability distribution function ({\small N-PDF})}\\
Shown on the panels of Fig. 5 are respectively the {\small PDF}s for gas column density (hereafter referred to as {\small N-PDF}s), at different epochs, for each choice of the magnitude of external pressure, $P_{ext}$, and the average initial density, $\bar{\mathrm{n}}$. Following extensive numerical work, it is now well established that gas dominated by turbulence is characterised by a lognormal distribution and develops a power-law extension towards higher column densities once self-gravity becomes dominant (e.g. Padoan \emph{et al.} 1997; Federrath \emph{et al.} 2008; Stanchev \emph{et al.} 2015). Let us begin by examining the {\small N-PDF}s for realisations 1, 2 and 3,  where the individual pre-collision flows had a density of $\bar{\mathrm{n}}=1$ cm$^{-3}$ and a relatively low magnitude of external pressure, $P_{ext}/k_{B}\sim 10^{3}-10^{4}$ K cm$^{-3}$, in Fig. 5.  Evidently, the collision between the initially warm atomic flows in each of these three cases appears to have culminated in nothing more than a nebula of diffuse HI. Albeit, the nebula in the third case, at least to some extent, appears to have some gas above the column-density threshold ($\gtrsim 10^{20}$ cm$^{-2}$; e.g. Stecher \& Williams 1967, Hollenbach \emph{et. al.} 1971, Federman \emph{et. al.} 1979), required for the formation of molecular hydrogen. The {\small N-PDF}s for these realisations appears to be a combination of lognormal distributions as has been shown by overlaying the constituent distributions on  the {\small N-PDF} for realisation 1. Thus with little gas in the dense phase, it appears that an environment of low interstellar pressure is likely inconducive to trigger star-formation. And if at all star-formation does eventually commence, it will probably be sluggish. \\ \\
We next examine the  {\small N-PDFs} for realisations 4, 5 and 6 in Fig. 5. The corresponding magnitude of external pressure, $P_{ext}$, as can be seen from Table 1, was an order of magnitude higher in comparison with that for the first three realisations. In contrast with the {\small N-PDF}s for realisations 1-3, these latter {\small N-PDF}s appear to be a combination  of lognormal-distribution at relatively lower densities and a power-law tail at higher densities that has a distinct break, especially at latter epochs.  That the power-law tail for these realisations has two components is readily visible for the respective {\small N-PDF}s. For illustrative purposes the constituent lognormal distribution at the low-density end of the distribution has been overlaid \textbf{with a dashed black line on the plot corresponding to realisation 4}. Also, the power-laws fitting the high-density end of the {\small N-PDF} for each realisation in this set has been shown on the individual plots. Please note, power-law fits have been shown for the {\small N-PDF} deduced at the latest epoch of the respective realisation. Thus, for instance, in the case of realisation 4, the power-law fits have been shown for the {\small N-PDF} corresponding to $t=4.2$ Myrs. Furthermore, the power-law fits shown for this realisation and those that follow later are good enough to be accepted at the 0.01 significance level of the simple Kolmogorov-Smirnov test. \\ \\
Evidently, with increasing magnitude of the external pressure, $P_{ext}$, both power-laws fitting the high-density end of the individual {\small N-PDF} become shallower or in other words, gas is cycled to increasingly higher densities. It is also interesting to note that the double power-law tail generated in this set of realisations (i.e. 4, 5 \& 6), is similar to the kind reported by Pokhrel \emph{et al.} (2016) for some regions within the cloud Mon R2.  Similarly, shown on the various panels of Fig. 6 are the {\small N-PDF}s for the remaining five realisations listed 7-11 in Table 1. As with the {\small N-PDF}s for the previous set of realisations (listed 4-6), the {\small N-PDF} for the post-collision slab in these realisations is also lognormal at early epochs  before subsequently developing a power-law tail at higher column densities. We note, in this set of realisations the {\small N-PDF} for realisations 7, 9 and 11 developed a double power-law tail at its high density end.  \\ \\
As before, the power-law fitting the high density end of the {\small N-PDF} has been shown for each realisation. In general, the {\small N-PDF}s for $P_{ext}\lesssim 2.56\times 10^{7}$ K cm$^{-3}$ (i.e., realisations 7-10), develop tails at their high-density end with relatively shallow slopes and the slope of this power-law is -0.6 for realisation 10.
In the somewhat extreme case such as in realisation 11, however, where $P_{ext}/k_{B}\sim 10^{8}$ K cm$^{-3}$, the {\small N-PDF} exhibits considerable steepening at the high density end; see lower panel of Fig. 6. We will revisit this point in \S 4.  We also note, the power-law extension in these latter realisations develops over a relatively shorter timescale in comparison with that for the earlier three realisations. \\ \\ \\
%---------------------------------------------------------------------
\begin{figure*}
\label{Figure 6}
\vspace{1pc}
\centering
\includegraphics[angle=270,width=0.45\textwidth]{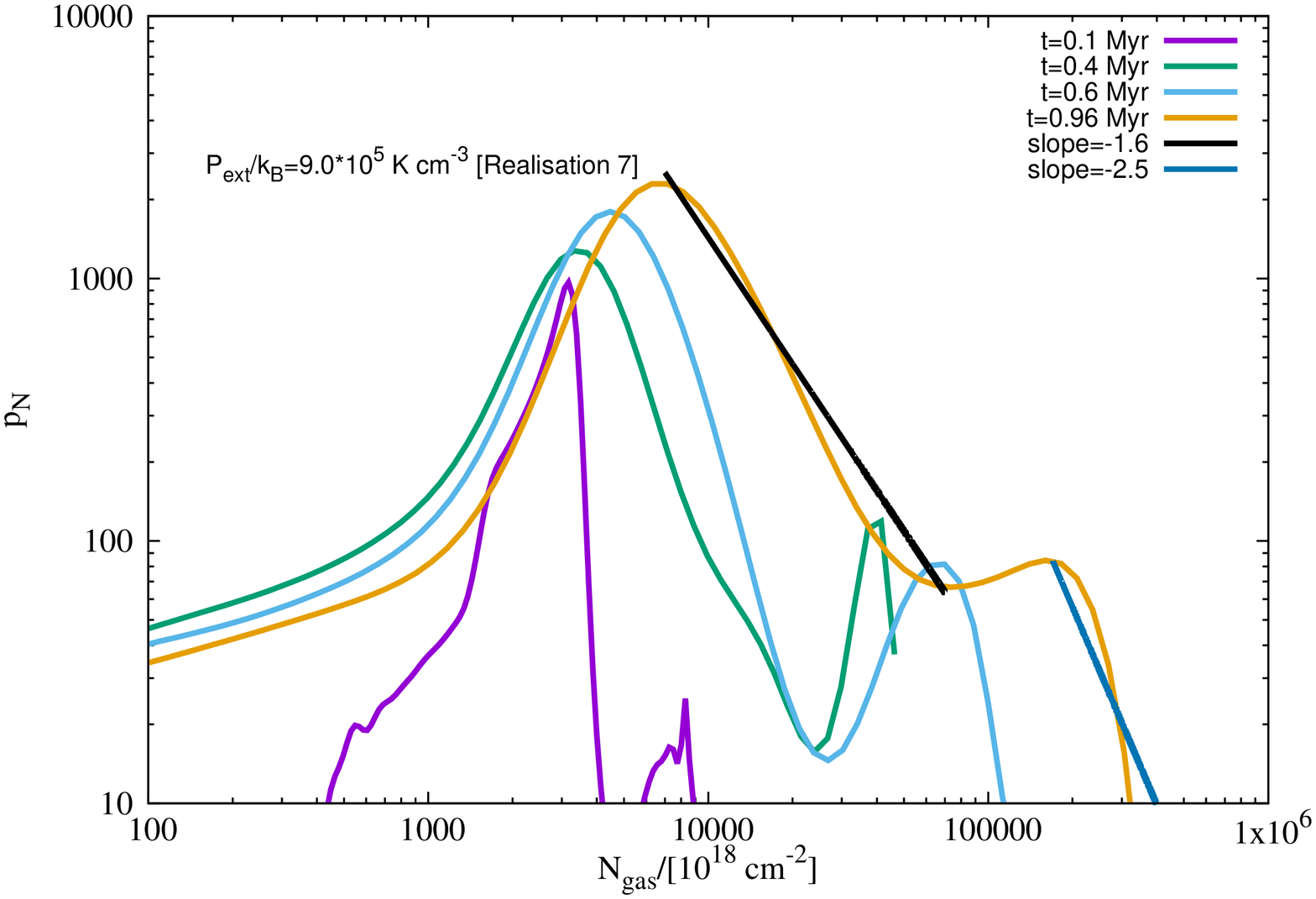}
\includegraphics[angle=270,width=0.45\textwidth]{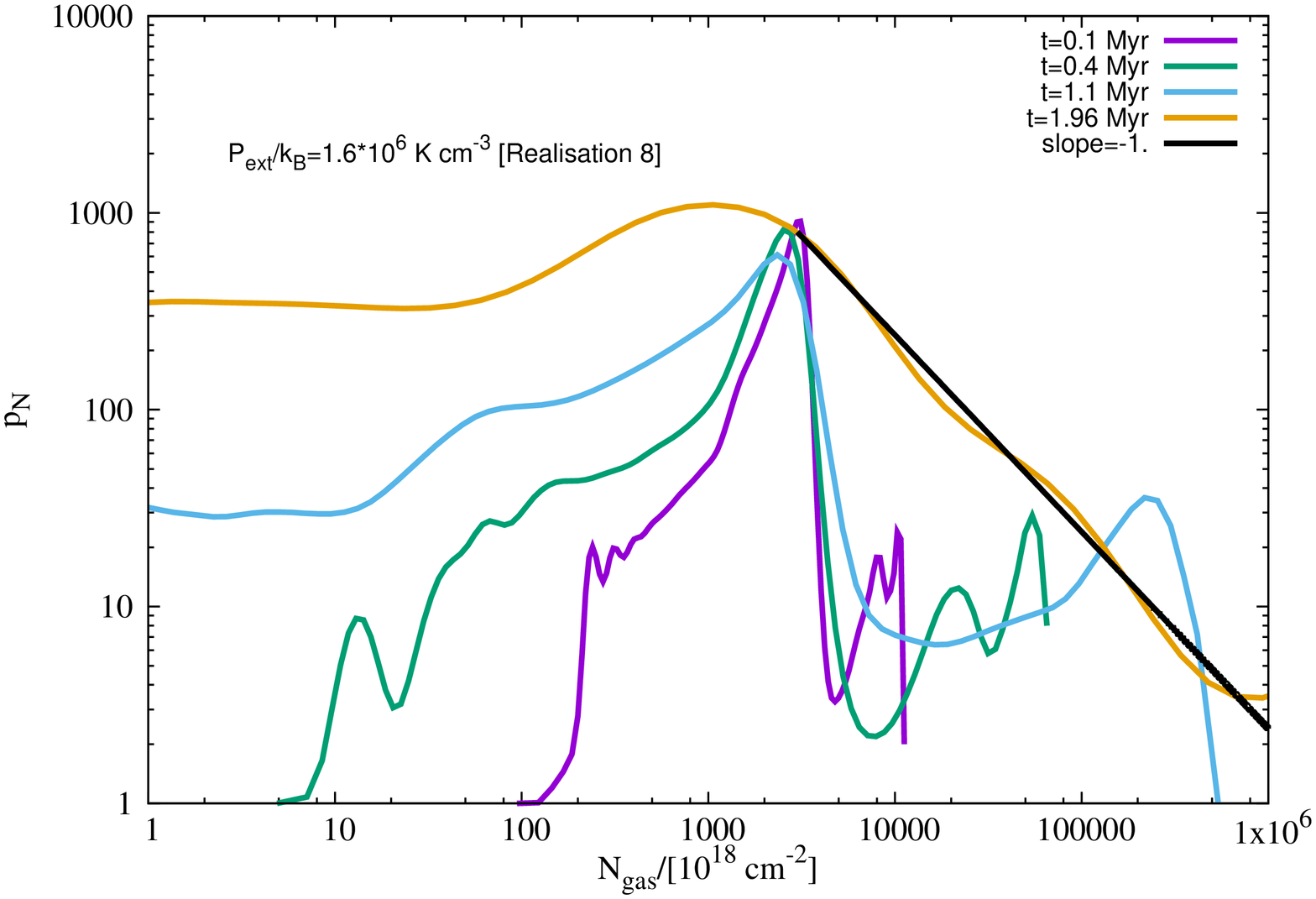}
\includegraphics[angle=270,width=0.45\textwidth]{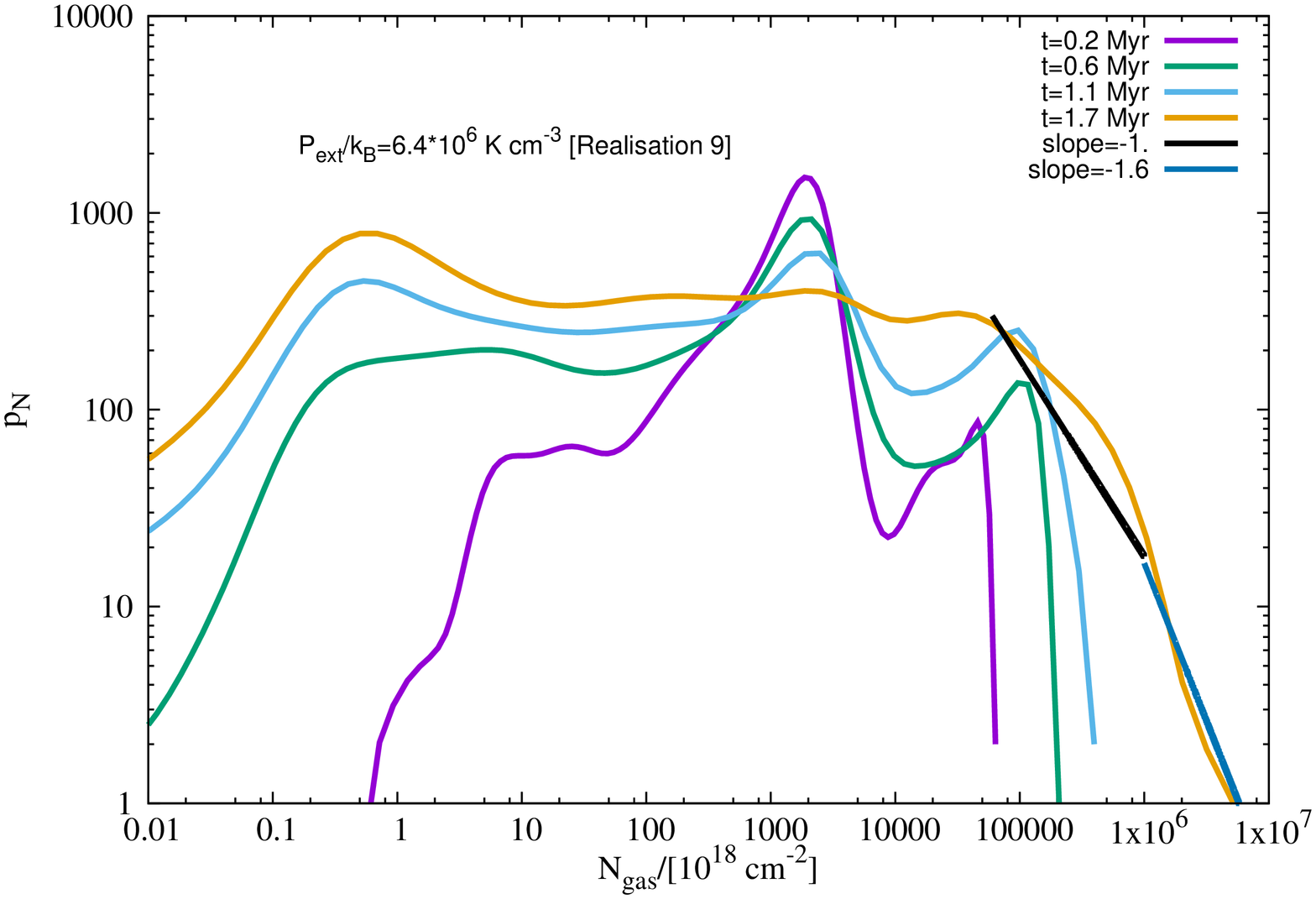}
\includegraphics[angle=270,width=0.45\textwidth]{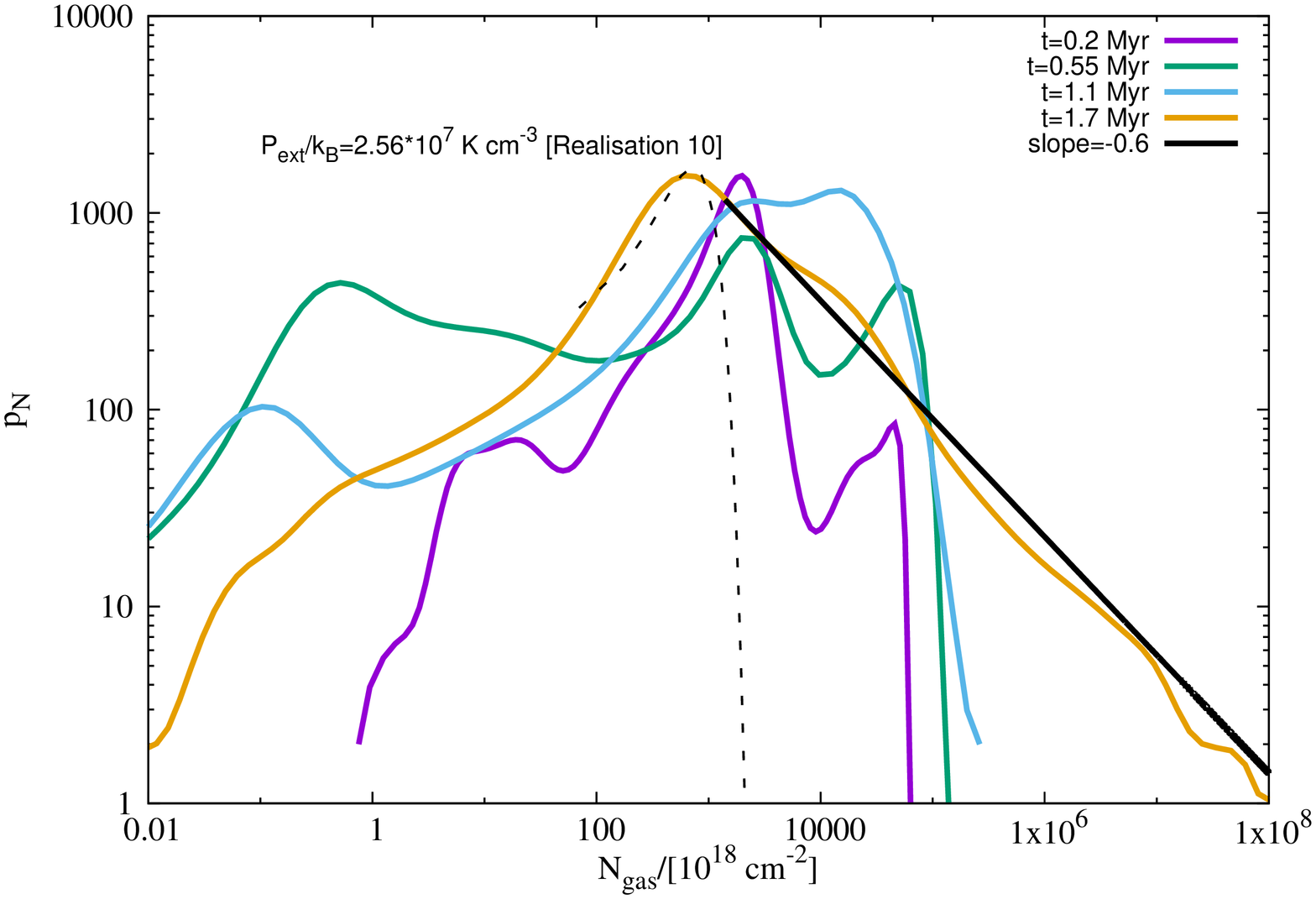}
\includegraphics[angle=270,width=0.45\textwidth]{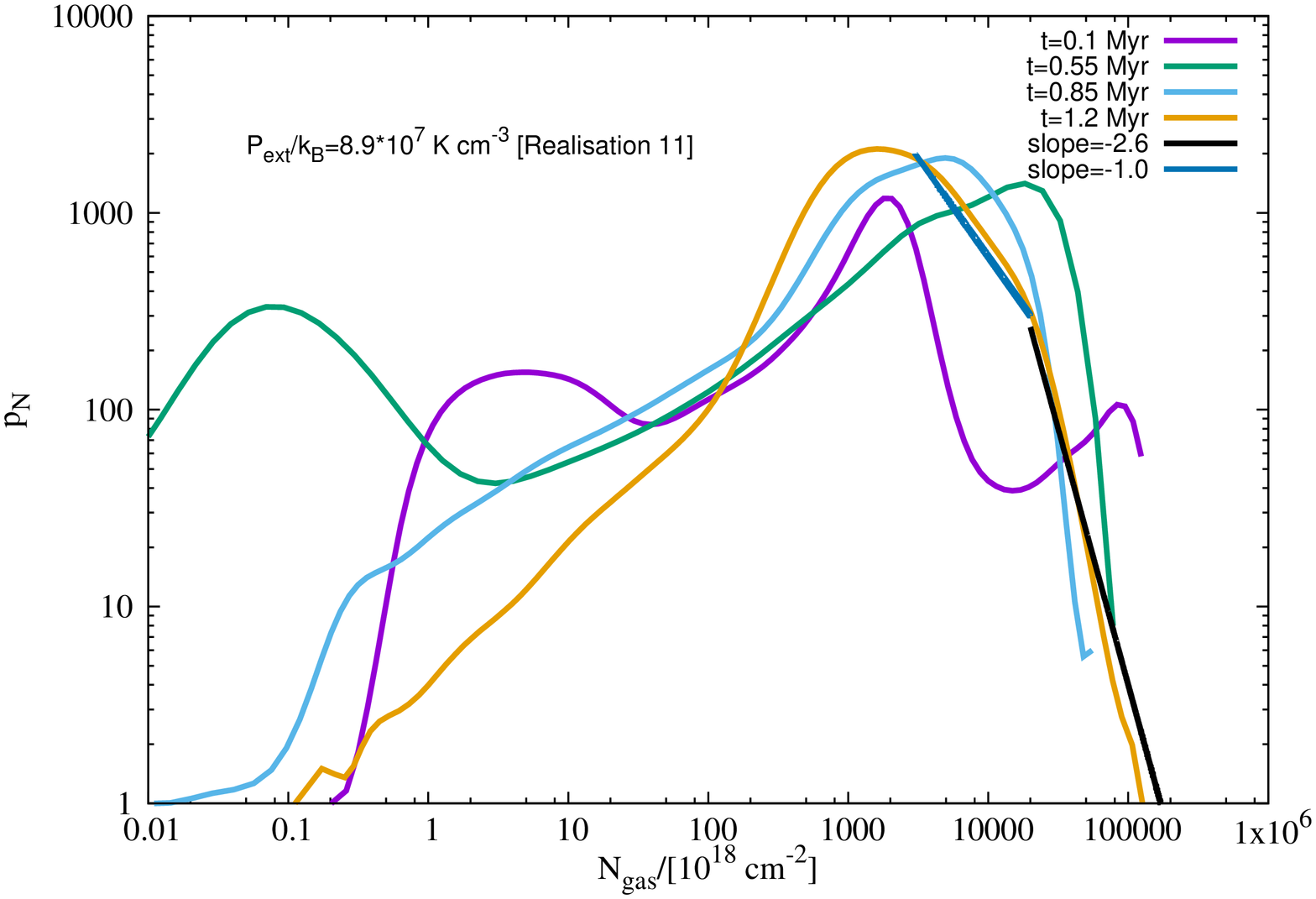}
\caption{Same as the plots in Fig. 5, but now for realisations listed 7-11 in Table 1.}
\end{figure*}
%
%--------------------------------------------------------------------
\begin{figure}[h]
\label{Figure 7}
\vspace{1pc}
\centering
\includegraphics[angle=270,width=0.5\textwidth]{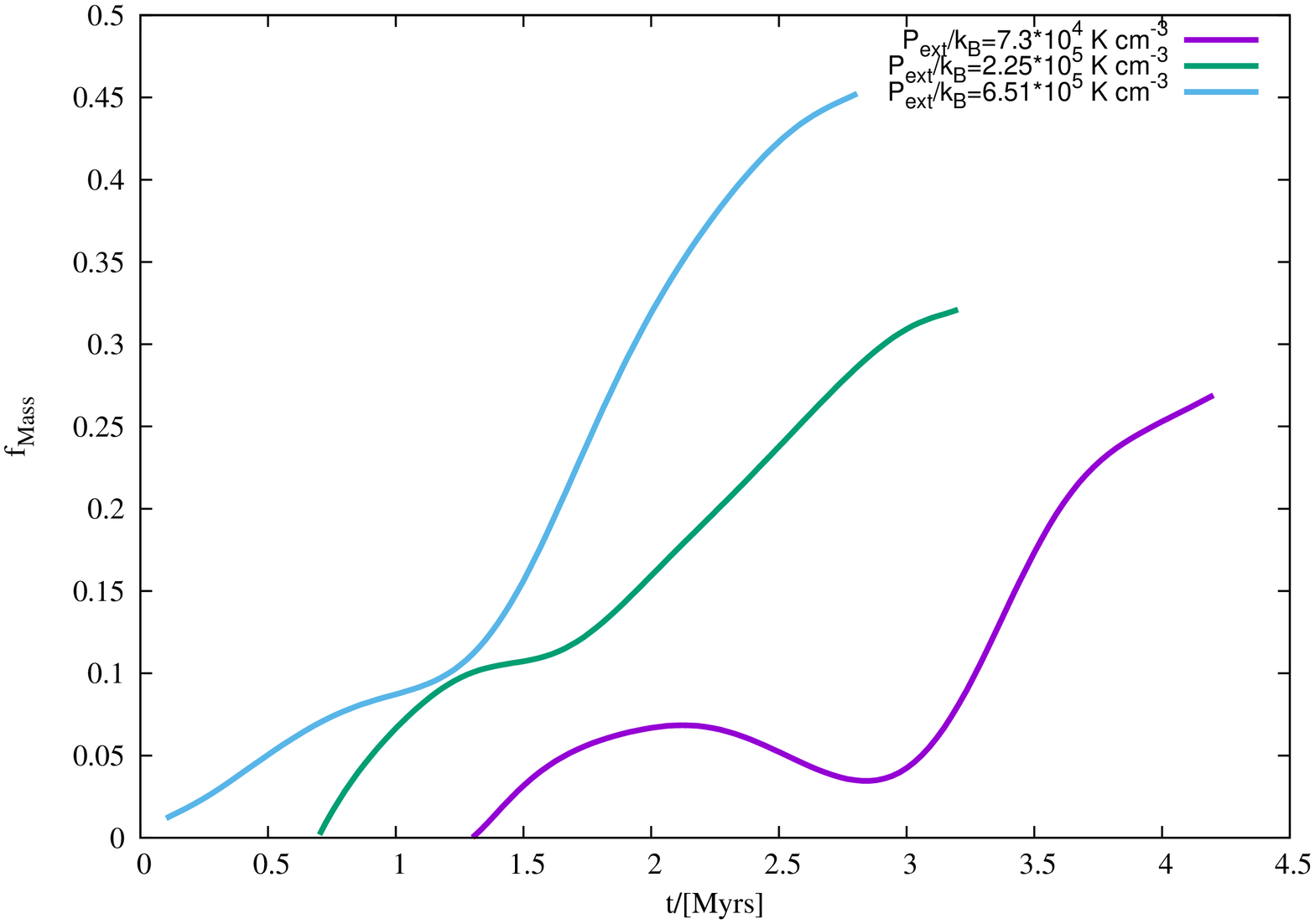}
\includegraphics[angle=270,width=0.5\textwidth]{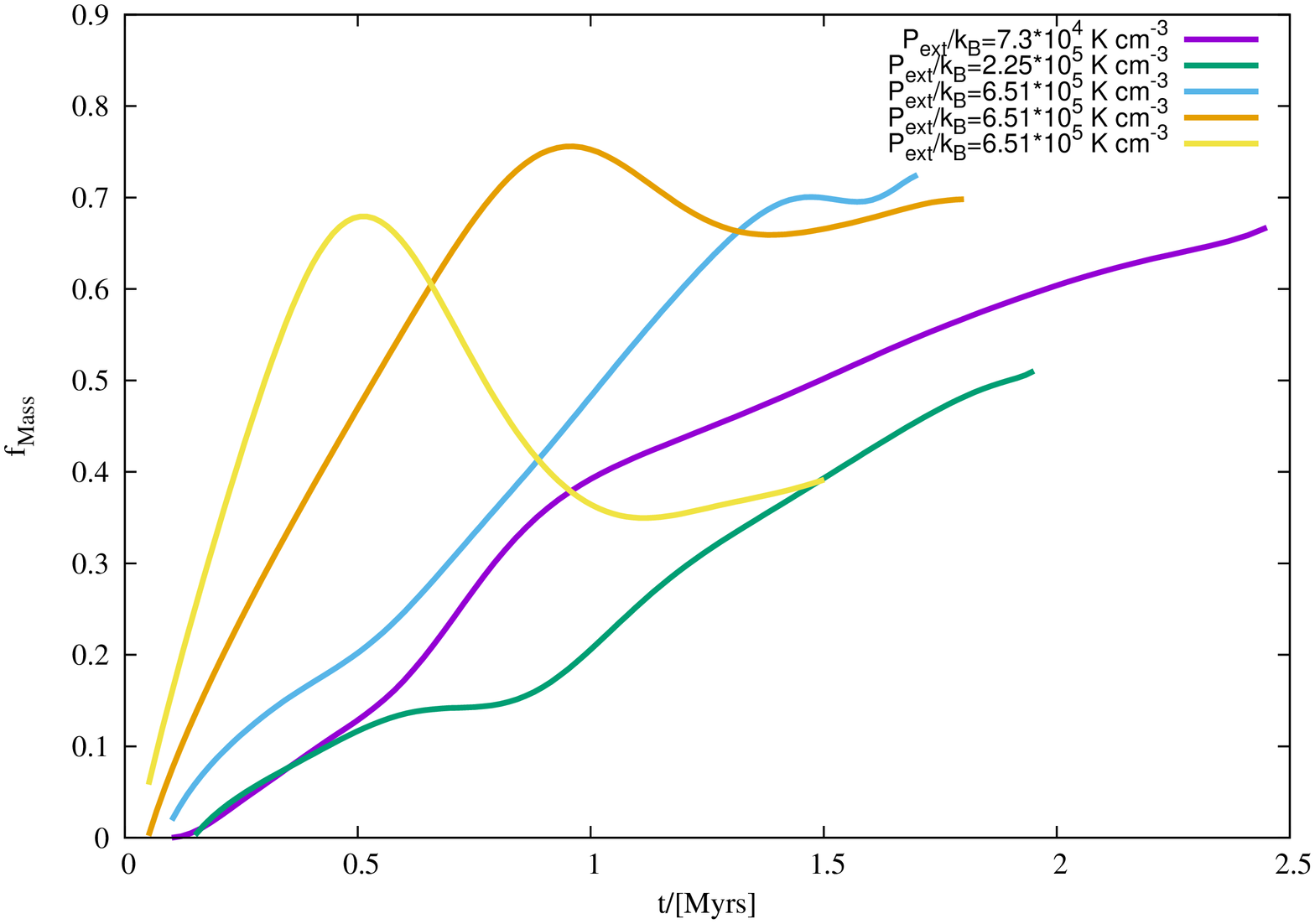}
\includegraphics[angle=270,width=0.5\textwidth]{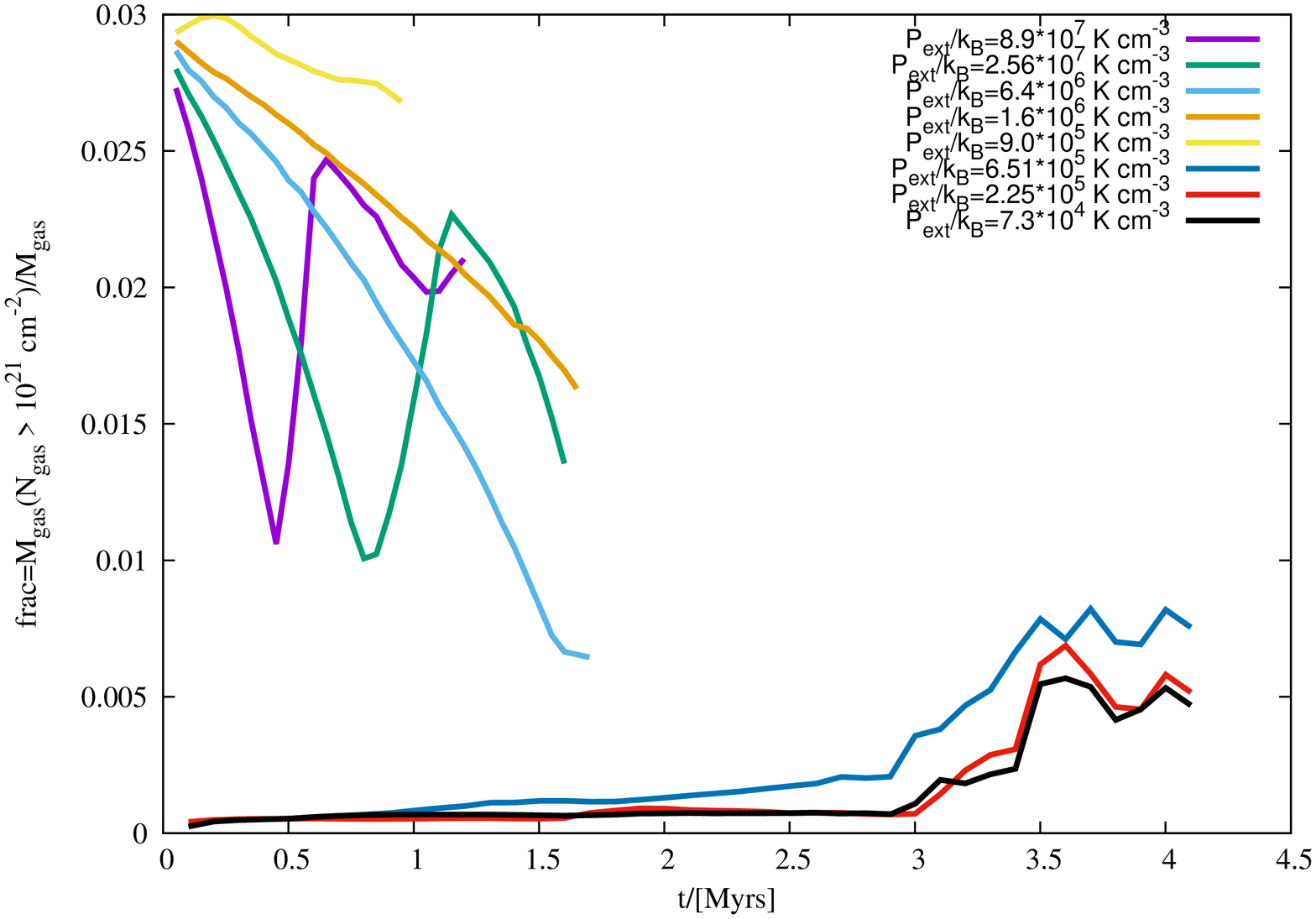}
\caption{Plots showing the temporal variation of the fractional mass of cold, dense gas; \emph{Upper-panel} : realisations 4-6, \emph{Central-panel} : realisations 7-11. Observe that the gas-depletion timescale becomes progressively smaller with increasing magnitude of $P_{ext}/k_{B}$; \emph{Lower-panel} : plot showing temporal variation of fractional mass of gas having column density greater than 10$^{21}$ cm$^{-2}$ in realisations 6-11.}
\end{figure}
%-------------------------------------------------------------------
\textbf{Dense gas-fraction \Big($f_{Mass}\equiv\frac{\mathrm{M}_{thresh}}{\mathrm{M}_{gas}}$\Big)} \\
Presently by dense, cold gas we imply the volume of putative star-forming gas and adopt a working threshold of gas denser than $\sim 10^{3}$ cm$^{-3}$ and colder than $\sim 50$ K.
This is the typical density of a potential star-forming clump though denser regions within such clumps have been mapped with the aid of the emission due to transitions of molecules such as {\small HCN}, {\small HCO}$^{+}$ and {\small HNC}. These emissions trace regions having density upwards of $\sim$10$^{4}$ cm$^{-3}$ (e.g. Gao \& Solomon 2004, Usero \emph{et al.} 2015 and Bigiel \emph{et al.} 2016). While such a steep density threshold could, in principal, have been adopted here to investigate the variation of the dense gas fraction, it would have picked only the more strongly self-gravitating pockets in the post-collision slab. Instead, we are presently interested in studying the variation in the fraction of gas potentially available for core-formation and have therefore adopted a threshold that is an order of magnitude lower, but consistent with that used by Ragan \emph{et al.} (2016) to quantify the star-forming fraction in the Galactic disk. \\ \\
 The first three realisations are obviously eliminated from this exercise, for the post-collision gas body in those respective cases remained diffuse and therefore warm at the time of terminating calculations. Shown on the upper and the central panel of Fig. 7 are respective plots showing the temporal evolution of this fraction in realisations 4-6, and 7-11. In general, the more energetic a collision, or larger the inflow velocity, $V_{inf}$, the higher the fraction of gas in the post-collision slab cycled into the dense, cold phase and shorter is the timescale of this cycling i.e., the timescale of gas-depletion in the cloud reduces progressively for an increasing magnitude of the external pressure. For the largest magnitude of external pressure, $P_{ext}/k_{B}\sim 10^{8}$ K cm$^{-3}$, up to 80\% of the gas appears in the dense, cold phase which in our realisation translates into a mass typically $\sim 9\times 10^{4}$ M$_{\odot}$, which is within a factor of two of the mass of the so-called 'Brick' in the Galactic {\small CMZ} (e.g. Longmore \emph{et. al.} 2014). \\ \\
There are, however, a few differences in the characteristics of this fraction as a function of external pressure. At intermediate magnitudes of pressure such as, $P_{ext}/k_{B}\sim (10^{5} - 10^{6})$ K cm$^{-3}$, the fraction, $f_{Mass}$, appears to gradually asymptote to $\sim 50\%$.  By contrast, for larger magnitudes such as, $P_{ext}/k_{B}\gtrsim$ 10$^{6}$ K cm$^{-3}$, $f_{Mass}$ attains a peak before eventually petering-out.
We attribute this difference to that in the difference in the strength of growth of the {\small NTSI}. As discussed in \S 3.1 earlier, the {\small NTSI} grows more vigorously in cases where the inflows are highly supersonic which presently is the case with realisations 10 and 11. The growth of this instability in the shocked-slab, as noted above, is associated with its rapid fragmentation and a strong shearing interaction between slab-layers. The observed decline in the dense-mass fraction in these latter realisations with a higher inflow velocity is the result of the {\small NTSI}-induced mixing between slab-layers. Then as dense filaments begin to form in the collapsed globule in realisation 10, the dense-gas fraction starts rising again. Note, however, this fraction for realisation 11 was still comparatively smaller at the time calculations were terminated. \\ \\
%--------------------------------------------------------------------
%
%--------------------------------------------------------------------
The difference in these characteristics becomes clearer from the plot shown on the lower-panel of Fig. 7. Shown in this plot is the temporal variation of the mass of gas having column density in excess of 10$^{21}$ cm$^{-2}$ in realisations 6-11. At intermediate magnitudes of pressure i.e., ($10^{5}\lesssim (P_{ext}/k_{B})\lesssim 10^{6}$) K cm$^{-3}$, this fraction increases steadily. At larger magnitudes of pressure i.e., ($10^{6}\lesssim (P_{ext}/k_{B})\lesssim 10^{7}$) K cm$^{-3}$, where the collision is relatively stronger and initially gas is quickly transferred into the dense-phase, but thereafter, as the {\small NTSI} dominates the evolution of the post-collision slab, this fraction gradually declines. Finally, in the extreme case where $(P_{ext}/k_{B})\gtrsim 10^{7}$ K cm$^{-3}$, this fraction oscillates between $\sim$1\% - 3\% which is due to the stronger {\small NTSI}-induced mixing between slab-layers and buckling of the slab-surface that rapidly creates and ruptures pockets of dense gas. Consequently, only a tiny fraction of gas is assembled into pockets with potential to spawn stars. \\
%---------------------------------------------------------------------
%---------------------------------------------
%-----------------------------------------------------------------------
%
\subsection{Physical properties of the assembled cloud}
So far having examined the physical properties of gas in the post-collision slab, we now examine the typical physical properties of this composite gas-body. For the purpose, we neglect any gas particles having density  smaller than $\sim 50$ cm$^{-3}$, the lower density threshold for typical Galactic clouds. The object so identified will hereafter be referred to as a cloud. The mass, $M_{cld}$, size, $L$, and the  surface density, $\Sigma_{gas}$, of this cloud was calculated as follows -
\begin{align}
M_{cld} &= \sum_{i} m_{i}, \\
      L &= 2\cdot\mathrm{max}<\textbf{r}_{c} - \textbf{r}_{i}>, \mathrm{and} \\
 \Sigma_{gas} &= \frac{M_{cld}}{L^{2}},
\end{align}
$m_{i}$, being the mass of individual particles, $i$, comprising this cloud; $\textbf{r}_{c}$, and $\textbf{r}_{i}$ respectively the centre of this cloud, and position of particles in this cloud. As noted in the Introduction,  the observationally deduced magnitude of average density for clouds suffers from biases introduced by the specific choice of gas-tracers. The average density of clouds ($\sim$100 cm$^{-3}$), deduced by for instance, Solomon \emph{et. al.} (1987), is largely based on CO-surveys which is a good tracer of relatively dense agglomerations of gas. A similar threshold has been used by several authors to identify clouds in their respective simulations (e.g. Fujimoto \emph{et al.} 2014). However, as has been discussed in \S 3.1 above, growth of the {\small NTSI} and the thermal instability in the post-collision slab segregate the dense and the rarefied phase in the post-collision slab. As a consequence, raising the density threshold for cloud-identification to one that would be comparable with the observationally deduced magnitude could, to some extent, blight the variations in the physical properties of the clouds assembled in our realisations. \\ \\
%-----------------------------------------------------
%----------------------------------------------------------
\textbf{Surface density of the post-collision cloud} \\
Shown on the upper left-hand panel of Fig. 8 is the temporal variation of the surface density, $\Sigma_{gas}$, of this cloud in four of our realisations. Here we use two realisations each as representative cases of the set of realisations corresponding to the choices of the initial density viz., $\bar{n}$ = 50 cm$^{-3}$ and 100 cm$^{-3}$. First, evident from this plot is a distinct trend where the peak magnitude of surface-density increases with increasing magnitude of external pressure, $P_{ext}$. Shown on the neighbouring plot on the right-hand panel of this figure is the surface-density of the cloud at different epochs in a realisation as a function of the magnitude of external pressure for that realisation. Data points marked with \textbf{green crosses} are those that correspond to the peak magnitude of the surface density for the respective realisation.
This plot has  been made for the same four realisations as before and the bold, black line signifying a proportionality between $\Sigma_{gas}$ and $P_{ext}$ appears to fit the \textbf{green crosses} reasonably well. Second, as with the dense gas fraction, $f_{Mass}$, the cloud surface density, $\Sigma_{gas}$, for realisations 10 and 11, as is visible in the plot on the left-hand panel of Fig. 8, also behaves differently in comparison with that for realisations 6 and 8 where the  magnitude of $P_{ext}$ was relatively lower. Thus while $\Sigma_{gas}$ tends to asymptote in the latter two realisations, in the former two, however, it achieves a peak magnitude before eventually falling-off as the cloud buckles strongly and finally collapses as the {\small NTSI} saturates at the latter stages of its growth.  \\ \\
This can be seen in the pictures on the lower-panel of Fig. 8. Shown here are the rendered density images of the shocked slab in realisation 10 at an early epoch ($t\sim 0.65$ Myr), when the slab was buckling with the {\small NTSI} in attendance and then at a later epoch ($t\sim 1.8$ Myrs), when it had collapsed to form an elongated globule. This picture is also interesting because it shows not only the formation of a network of filaments, but also two other more prominent filaments that are likely to collide close to the centre of this globule. Colliding and/or interacting filaments have also been reported in recent literature. For instance, according to Nakamura \emph{et al.} (2014), star-formation in the Serpens South was likely triggered by a collision between filamentary clouds. Similarly, the filaments in the Cygnus {\small OB7} {\small MC} appear to be colliding and which according to numerical simulations developed by Dobashi \emph{et al.} (2014), cannot possibly be reconciled without invoking an external velocity gradient in the host cloud. The picture in the lower right-hand panel of Fig. 8 is consistent with the argument presented by these authors. The required velocity gradient in this realisation is invoked naturally by virtue of the collapse of the post-collision slab. Furthermore, there is also evidence of colliding 
filaments triggering the formation of dense clumps in the Perseus {\small MC} (e.g. Frau \emph{et al.} 2015).  \\ \\
Earlier we noted that the surface density for realisation 6 appeared to grow asymptotically unlike that for cases 10 and 11. It must be remembered that realisation 6 was developed with a relatively lower magnitude of the inflow velocity, $V_{inf}$, and therefore a proportionally smaller magnitude of external pressure, $P_{ext}$. Now, although the evolution of the slab in this case was qualitatively similar to that in realisations 10 or 11, and as is visible from the rendered images shown in Fig. 2, the {\small NTSI} in this case grew on a relatively longer timescale. The picture on the right-hand panel of Fig. 2 shows that at $t\sim 2.81$ Myrs the growth of this instability was yet to acquire its saturation. At this epoch,  before the slab could collapse, the realisation in this case had been terminated. As remarked earlier, the {\small NTSI} grows faster for higher inflow velocities so that in the present work it grew the fastest in realisations 10 and 11, those with the highest inflow velocity.  \\ \\
%---------------------------------------------------------------
\begin{figure*}
\label{Figure 8}
\vspace{0pc}
\centering
\includegraphics[angle=270,width=0.45\textwidth]{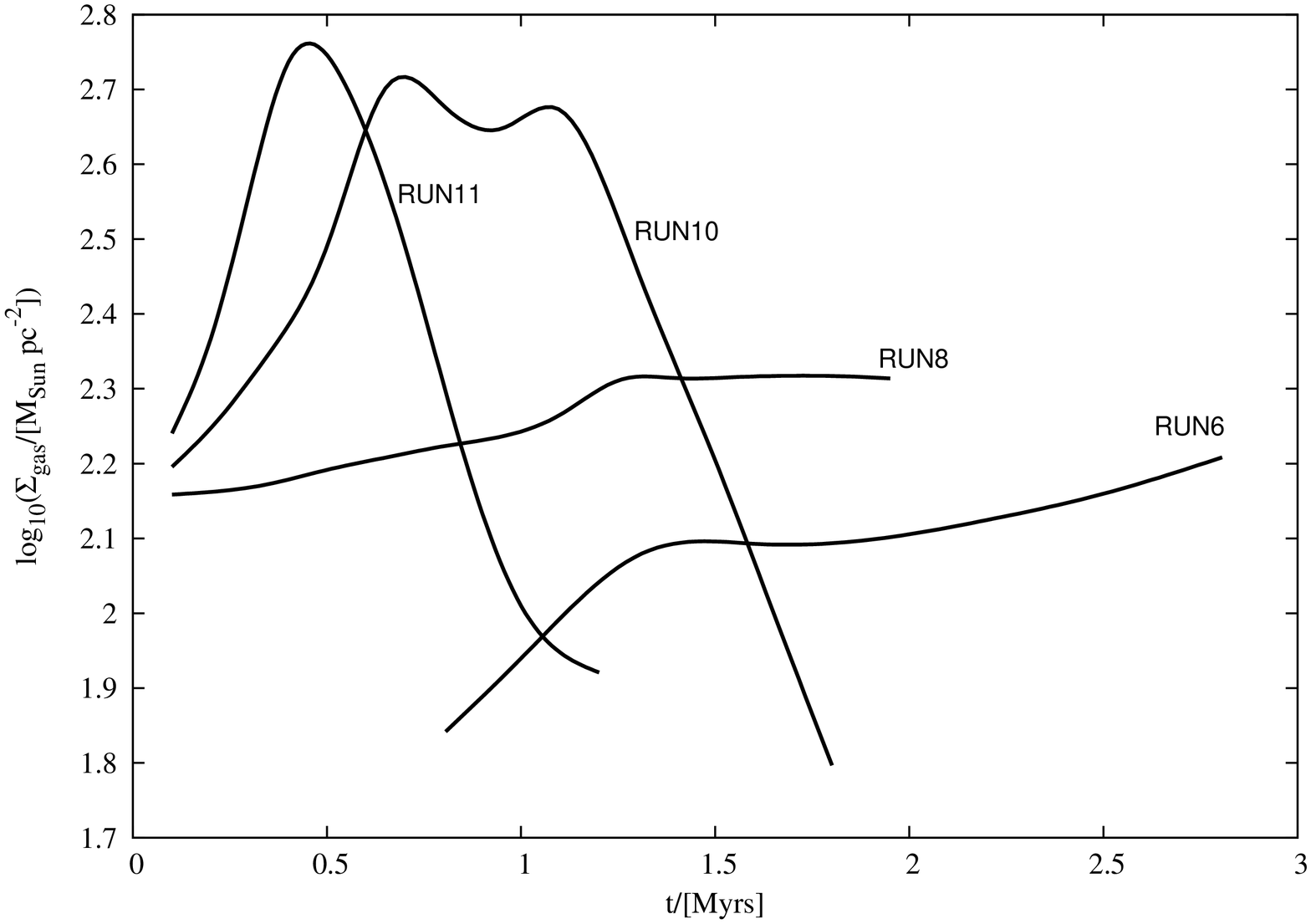}
\includegraphics[angle=270,width=0.45\textwidth]{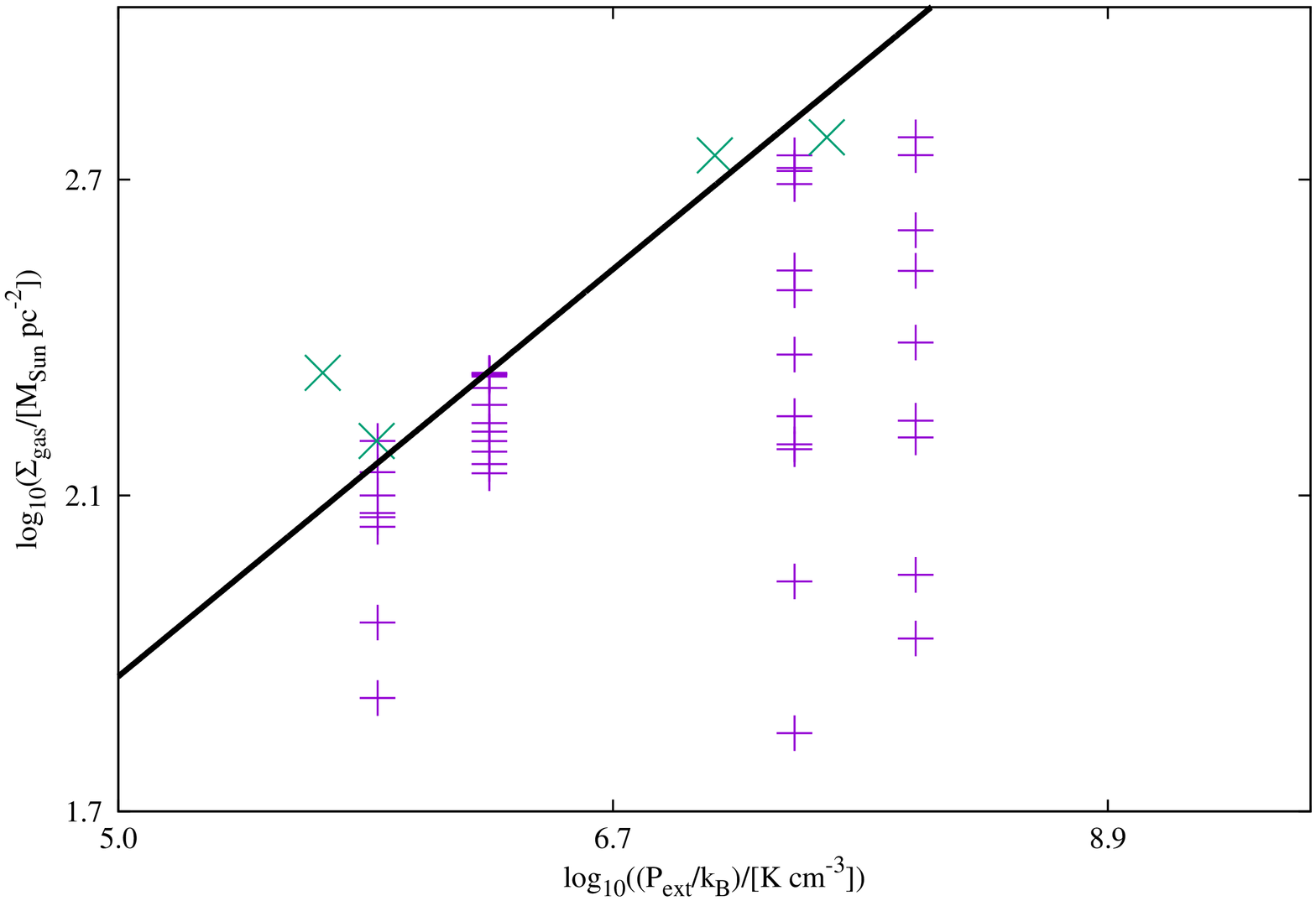}
\includegraphics[angle=270,width=\textwidth]{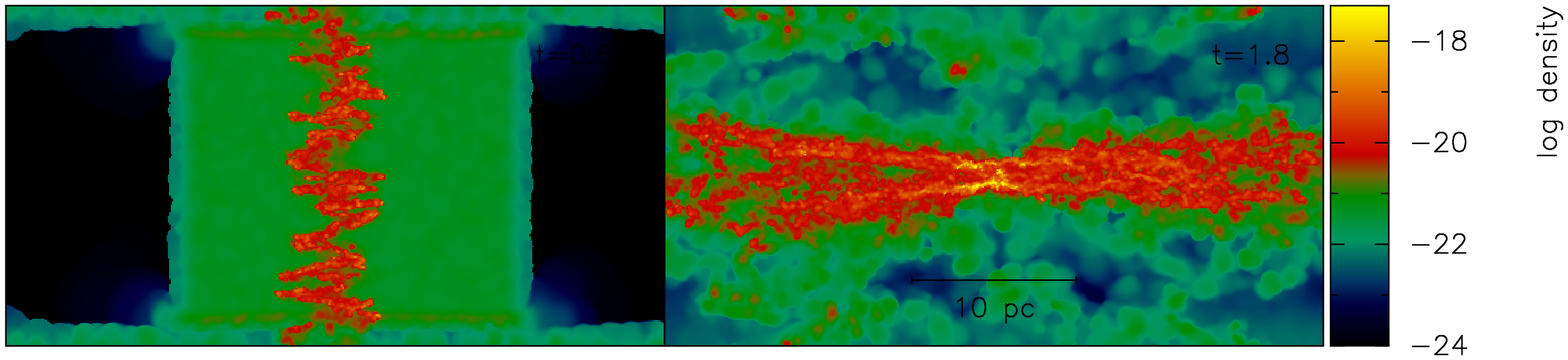}
\caption{\emph{Upper left-hand panel} : Temporal variation of the surface density, $\Sigma_{gas}$, of the cloud in each of the four representative realisations; \emph{Upper right-hand panel} : Surface density, $\Sigma_{gas}$, of the cloud in these four realisations at different epochs of its evolution as a function of the corresponding external pressure, $P_{ext}/k_{B}$. The bold line represents direct proportionality between $\Sigma_{gas}$ and $P_{ext}/k_{B}$. \emph{Lower left-hand panel} : Rendered density image showing the slab in realisation 10 at an earlier  epoch ($t\sim$0.65 Myr), when it had its peak surface density;
\emph{Lower right-hand panel} : Rendered density image of the same slab at a latter epoch ($t\sim$ 1.8 Myr), when its surface density had fallen significantly. At this latter epoch, the slab has collapsed to form an elongated globule and in which a plethora of thin filaments have started forming; close to the middle of the collapsed globule, two large filaments appear to be interacting.}
\end{figure*}

\begin{figure}
\label{Figure 9}
\vspace{0pc}
\centering
\includegraphics[angle=270,width=0.5\textwidth]{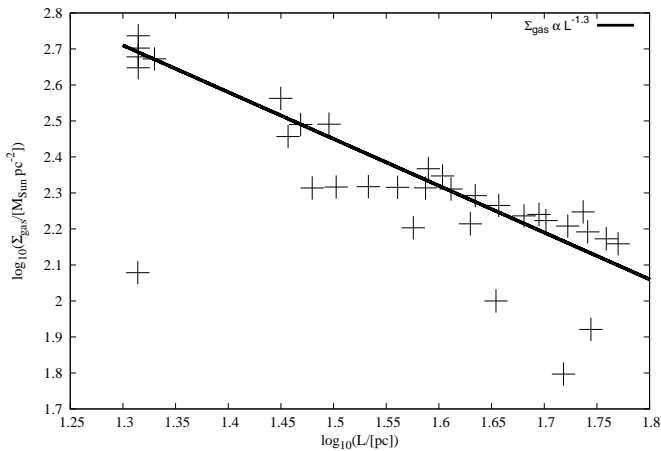}
\caption{Test of Larson's third-law : Shown in this plot is the surface density, $\Sigma_{gas}$, of the cloud as a function of its size at different epochs of its evolution.}
\end{figure}
%
%-----------------------------------------------
\textbf{Larson's Third Law : Do clouds indeed have uniform surface density ?} \\
We now come to the next important point, that about the validity of Larson's third law. An important conclusion that follows from the Larson's scaling relations (Larson 1981), is that clouds are entities having approximately uniform surface densities and specifically, the surface density varies weakly with cloud-size ($\propto L^{-0.1}$). We test this conclusion by generating a surface density-size plot for clouds at different epochs in the same four realisations as above. Shown in Fig. 9 is the resulting plot from which it is evident that the surface density of clouds is far from uniform and in fact, can be fit reasonably well with a power-law of the type $L^{-1.3}$. So, evidently the clouds simulated in this work are inconsistent with the Larson's third law. This is hardly surprising, for as shown above, physical dimensions of the assembled clouds undergo significant changes over the course of their evolution by virtue of the growth of the {\small NTSI}. \\ \\
%------------------------------------------------------------------
%
%---------------------------------------------------------------
\textbf{Internal pressure vis-a-vis the external pressure} \\
The calculation of the magnitude of internal pressure, $P_{int}/k_{B}$, for the cloud assembled in these realisations includes contribution from the thermal ($P_{th}/k_{B}\equiv \bar{n}_{gas}\bar{T}_{gas}$; $\bar{n}_{gas}$ and $T_{gas}$ are respectively the mean gas density, and temperature of the cloud), and the non-thermal component ($P_{NT}/k_{B}\equiv \mu\bar{n}_{gas}\sigma_{gas}^{2}/k_{B}; \mu$, being the mean molecular weight of the gas).
The corresponding plots on the left-hand panel of Fig. 10 show that like the surface density, the peak magnitude of the internal pressure also increases with increasing magnitude of the external pressure, $P_{ext}/k_{B}$. In each of these realisations the observed peak in $P_{int}/k_{B}$ was largely due to a higher contribution from the non-thermal component. Note also that the magnitude of internal pressure tends to rise spectacularly at later times for realisations 8 and 10. This observed increase in $P_{int}/k_{B}$ was primarily due to a higher contribution from the thermal component of the gas which in the collapsed globule in realisation 10 for instance, had become strongly self-gravitating. The dominance of self-gravity manifested itself in the form of dense filaments visible in the lower right-hand panel of Fig. 8. The calculations for realisation 11 were terminated soon after the shocked-slab consumed the inflows and began to exhibit strong buckling. It is for this reason that a similar sharp increase in $P_{int}/k_{B}$ is not visible in the characteristic for this latter realisation.  \\ \\
The plot on the right-hand panel of Fig. 10 demonstrates that the observed increase in the magnitude of internal  pressure, $P_{int}/k_{B}$, remains within about an order of magnitude of the external pressure, $P_{ext}/k_{B}$, for the respective realisation. At no instance does the cloud in any realisation ever become over-pressured, not even for a relatively large magnitude of external pressure, $P_{ext}/k_{B}\gtrsim 10^{7}$ K cm$^{-3}$. In all these  realisations the assembled cloud at best acquired a configuration where $P_{ext}/k_{B}\sim P_{int}/k_{B}$, i.e., it tended toward a configuration obeying the pressure-modified Virial equilibrium ({\small PVE}), and thereafter, it became pressure-confined. In fact, at the epoch when the respective clouds acquired their peak magnitude of $P_{int}$, marked with \textbf{green crosses} on the plot shown on  the right-hand panel of Fig. 10, which is also the epoch when the respective clouds had acquired their peak density, $\Sigma_{gas}$, as shown in Fig. 8 above, the clouds were in approximate pressure-equilibrium.
It therefore appears that after all molecular clouds may, at some stage of their evolution, appear to be obeying the {\small PVE}. Strictly speaking, it may not even be necessary to treat clouds as objects in pressure equilibrium, for they appear to be so only briefly during their evolutionary sequence; see also Heitsch \emph{et al.} (2008a) for a discussion on the matter. \\ \\
%-----------------------------------------------------------
\begin{figure*}
\label{Figure 10}
\vspace{0pc}
\centering
\includegraphics[angle=270,width=0.45\textwidth]{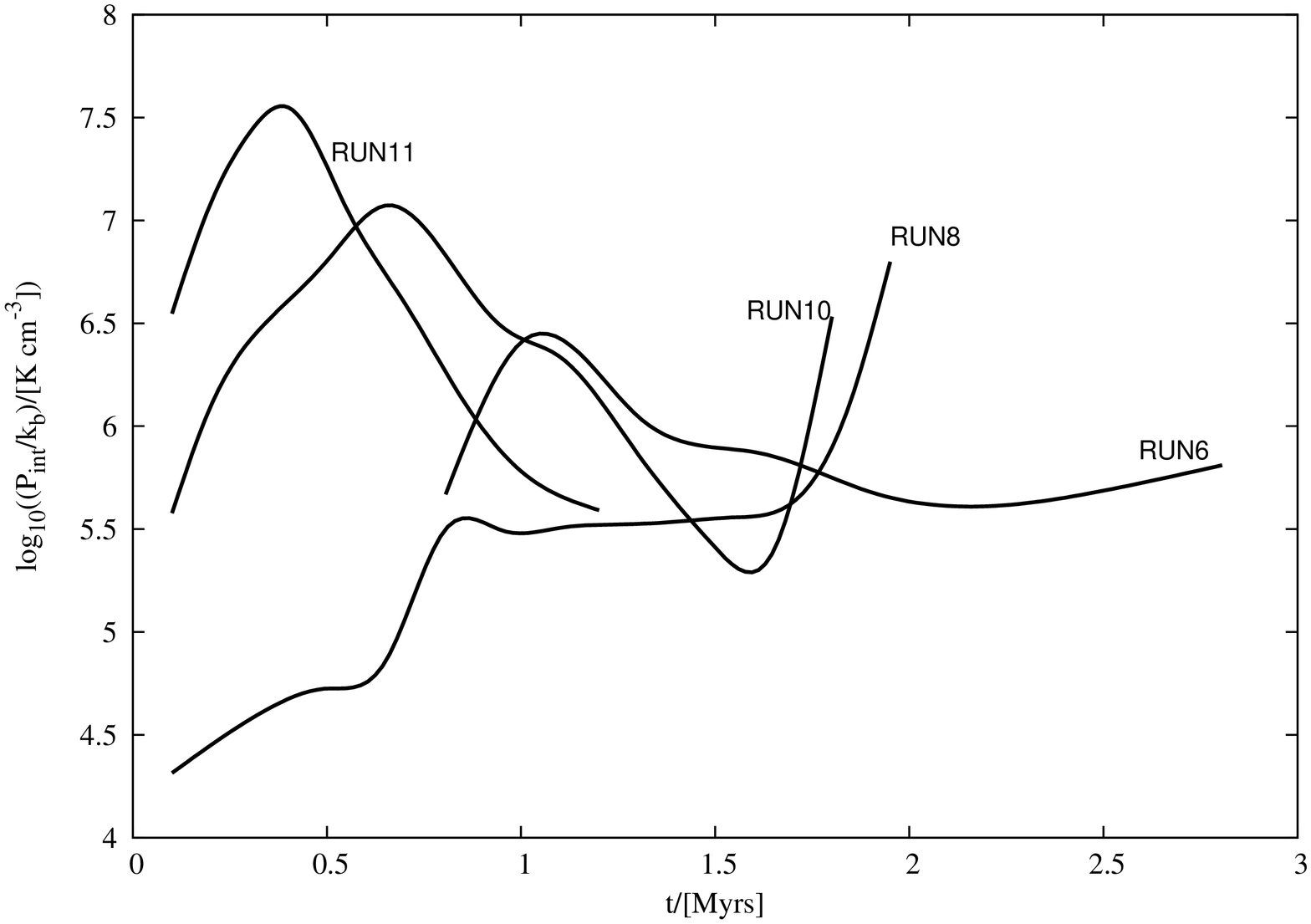}
\includegraphics[angle=270,width=0.45\textwidth]{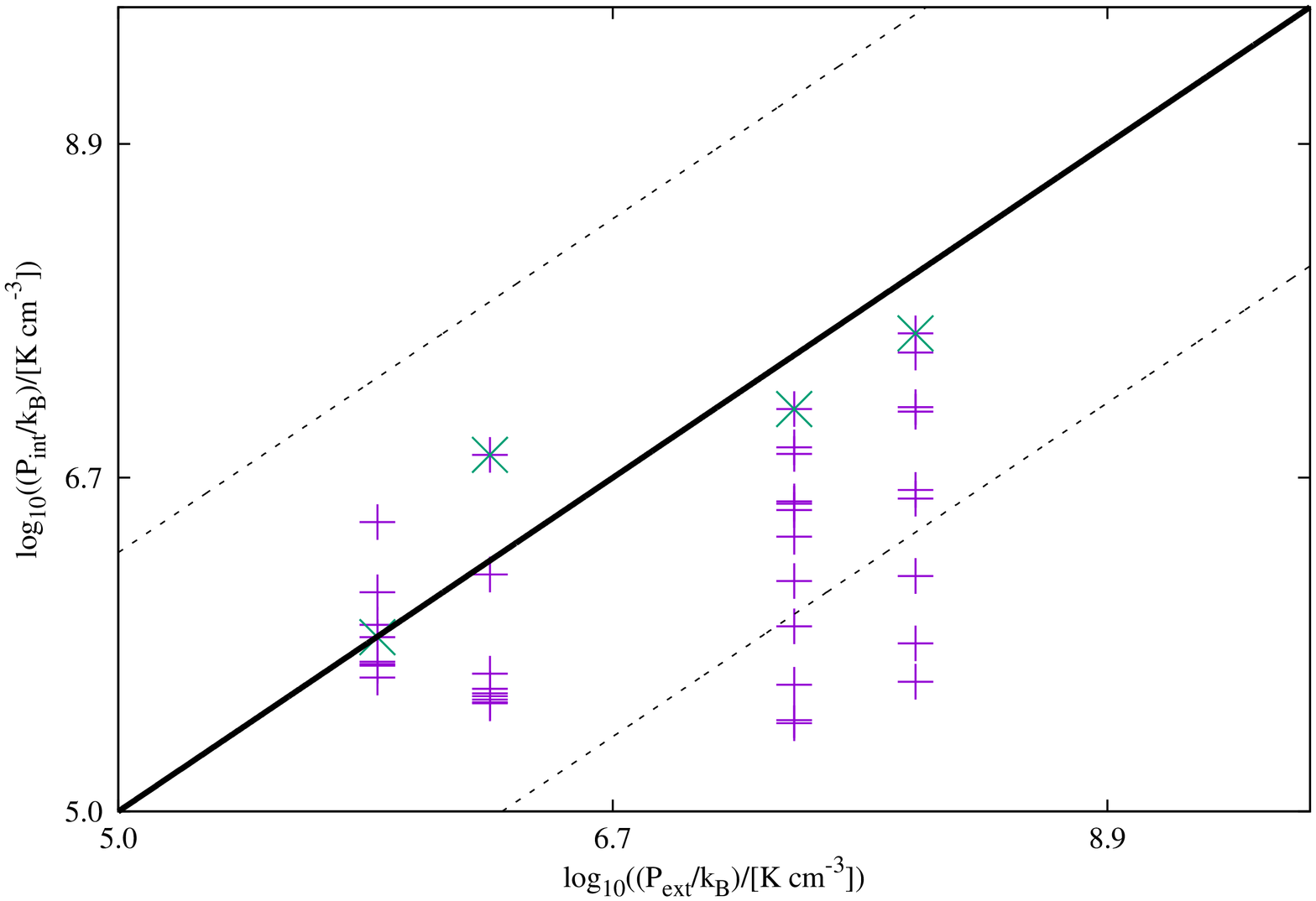}
\caption{\emph{Left-hand panel} : Shown in this plot is the temporal variation of the internal pressure, $P_{int}/k_{B}$, of the cloud for these four representative realisations. \emph{Right-hand panel} : The magnitude of internal pressure, $P_{int}/k_{B}$, at different epochs of the cloud in these four realisations as a function of the respective external pressure, $P_{ext}/k_{B}$. The bold black line corresponds to a direct proportionality between $P_{int}/k_{B}$ and $P_{ext}/k_{B}$; the dotted lines on either side of this line represent an order of magnitude variation in pressure.}
\end{figure*}
%------------------------------------------------------------
\textbf{Are Molecular clouds in Virial equilibrium ?} \\
Following the convention adopted by Field \emph{et al.} (2011), we wish to distinguish between the Simple Virial equilibrium ({\small SVE}), and the pressure modified Virial equilibrium ({\small PVE}). The latter accounts for the contribution due to the external pressure. The velocity dispersion, $\sigma_{gas}$, of gas in a cloud of size, $L$, is related to its column density, $\Sigma_{gas}$, and the external pressure according to -
\begin{equation}
\frac{\sigma_{gas}^{2}}{L} = \frac{1}{6}\Big(\pi\Gamma G\Sigma_{gas} + \frac{4P_{ext}}{\Sigma_{gas}}\Big)
\end{equation}
Field \emph{et. al.} (2011); $\Gamma$=0.6 for a cloud having uniform density. In the corresponding expression for the {\small SVE} there is no contribution due to the term from external pressure. The plots shown in Fig. 11 here are largely the same as those presented by Field \emph{et al.} (2011) in their Fig. 3. Overlaid on top of the V-shaped characteristics for different magnitudes of $P_{ext}$ is the data for clouds from Heyer \emph{et al.} (2009), as well as those for clouds from the {\small LMC} and the {\small SMC} adopted from Bolatto \emph{et al.} (2008). The bold black line represents the {\small SVE}. The distribution of these cloud data about the {\small PVE} characteristics was interpreted by Field \emph{et. al.} (2011) as evidence suggesting that the size-linewidth relation for clouds is modified by external pressure and in fact, its proportionality is not constant but dependant on the magnitude of external pressure. \\ \\
In order to examine if the clouds assembled in our simulations do obey the {\small SVE} or indeed the {\small PVE}, we calculated the ratio on the left-hand side of Eqn. (7), the square of the size-linewidth coefficient, for them and plotted them along with the solutions for the {\small PVE} and the Heyer \emph{et al.} data in Fig. 11. It can be readily seen that clouds in these  realisations do not obey the {\small SVE}. The size-linewidth coefficient is progressively higher for larger magnitudes of external pressure, $P_{ext}$, and that the characteristic for a single magnitude of $P_{ext}$ cannot reconcile the dynamical state of the assembled cloud. In other words, the cloud assembled in a realisation, as it evolves with time, cuts across the {\small PVE} solution for that magnitude of external pressure, $P_{ext}$. For instance, the cloud assembled in realisation 10 with $P_{ext}/k_{B}\sim 2.56\times 10^{7}$ K cm$^{-3}$ makes its initial appearance below the {\small PVE} solution for $P_{ext}/k_{B}$ = 10$^{6}$ K cm$^{-3}$, thereafter as it continues to accrete gas, it makes an excursion beyond the {\small PVE} solution for $P_{ext}/k_{B}$ = 10$^{7}$ K cm$^{-3}$ and eventually arrives at a point between the {\small PVE} solutions for 10$^{6}$ K cm$^{-3}$ and 10$^{7}$ K cm$^{-3}$ at a later epoch. \\ \\
Here we note that Eqn. (7) is a relatively simple expression that does not account for the complex dynamics of a shocked-slab and so, it overestimates the size-linewidth coefficient for a given magnitude of surface density, $\Sigma_{gas}$, and external pressure, $P_{ext}$. As has been shown above, both the gas velocity-dispersion and the surface density of the post-collision cloud changes continuously as the {\small NTSI} grows and attains saturation. Consequently, one sees an excursion of the post-collision cloud about the {\small PVE} characteristics shown in Fig. 11 for a given magnitude of $P_{ext}$. Thus there is only brief period when the cloud obeys the {\small PVE} as it intersects the {\small PVE} characteristic corresponding to the magnitude of $P_{ext}$ for a realisation. This was also reflected in the plots on the left-hand panel of Fig. 10 when briefly there was an approximate equality between $P_{int}$ and $P_{ext}$.
This observation leads us to the inference that we are presently observing field clouds at different stages of their evolution and which therefore occupy different locations on this plot. It also lends further credence to the hypothesis proposed by Field \emph{et al.} (2011) that the observed dynamical properties of clouds could possibly be reconciled with not a single magnitude of $P_{ext}$, but a range of magnitudes typically between 10$^{4}$ K cm$^{-3}$ and 10$^{8}$ K cm$^{-3}$.
%-------------------------------------------------------------------------
\begin{figure*}
\label{Figure 11}
\vspace{1pc}
\centering
\includegraphics[angle=270,width=\textwidth]{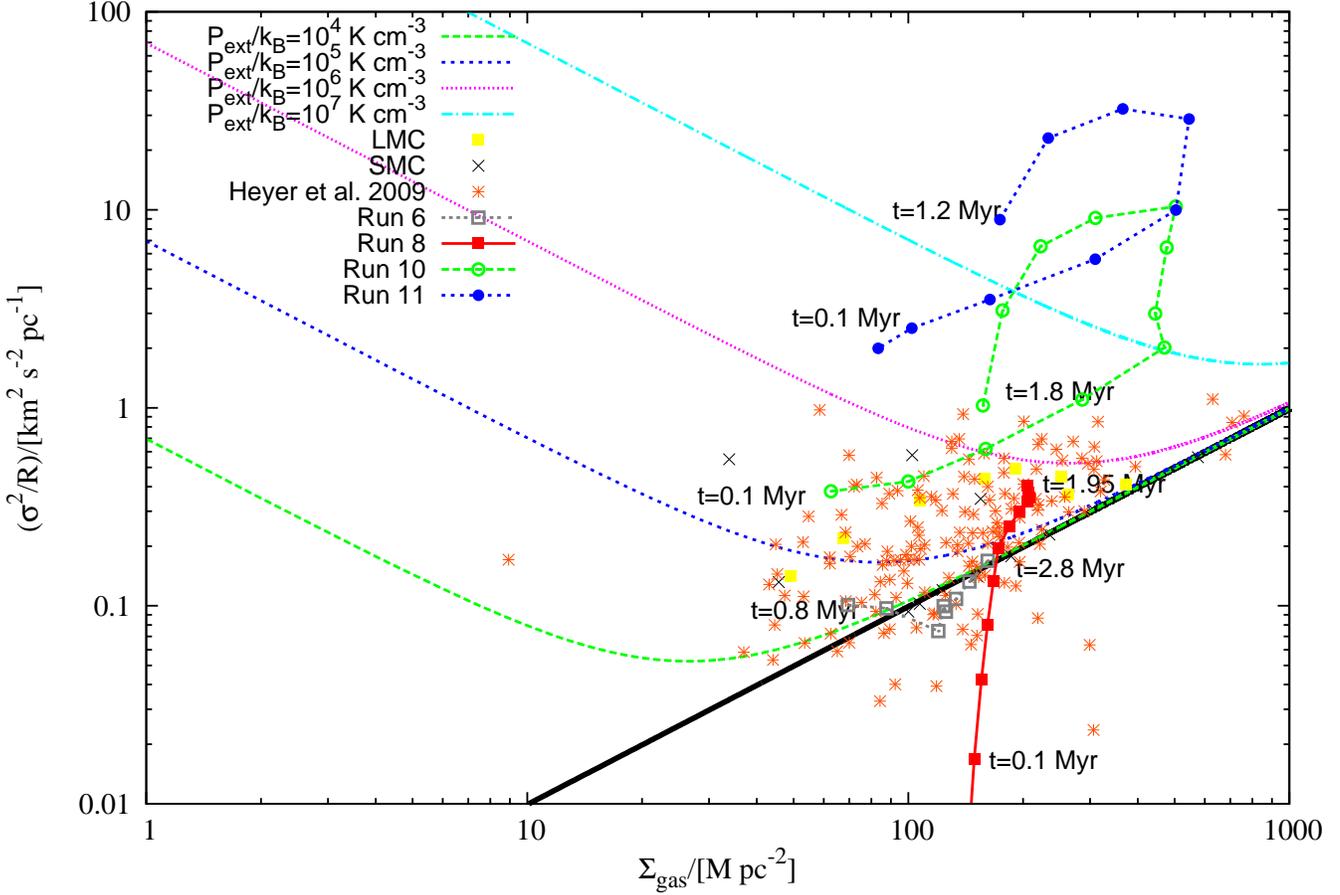}
\caption{The variation of the size-linewidth coefficient for a cloud with its surface density. This is the same plot as Fig. 3 in Field \emph{et al.} (2011), but now overlaid with the temporal locations of clouds in four of our realisations. The 'V'-shaped characteristics are the solutions of the {\small PVE} given by Eqn. 7 for specific magnitudes of $P_{ext}/k_{B}$.}
\end{figure*}
\section{Discussion}
There are largely two schools of thought as regards the nature of molecular clouds ({\small MC}s). On the one hand, Krumholz \& McKee (2005) argued that {\small MCs} are objects in Virial equilibrium while on the other, Ballesteros-Paredes (2006) and other authors argued that clouds are unlikely to be in Virial equilibrium, but rather objects which at best represent equipartition between self-gravity, the kinetic energy and the magnetic energy. See also the review by Hennebelle \& Falgarone (2012) for a further discussion of this point. The Krumholz-McKee hypothesis is supported by a number of older cloud surveys such as those by Solomon \emph{et al.} (1987), Rosolowsky \emph{et al.} (2003, 2005) and Bolatto \emph{et al.} (2008). However, this hypothesis is inconsistent with the more recent findings such as those reported by Heyer \emph{et al.} (2009), Hughes \emph{et al.} (2010, 2013b), Meidt \emph{et al.} (2013) and Rice \emph{et al.} (2016). \\ \\
These latter studies and in particular Heyer \emph{et al.} (2009) argue that cloud masses in a number of earlier surveys were significantly overestimated leading them to being classified as Virialised objects. These authors also reported a variation in the size-linewidth coefficient in proportion with the surface-density of clouds in their data sample. This is inconsistent with the principal conclusion of Larson (1981) that column density of clouds is approximately constant and as a consequence, the size-linewidth coefficient must not vary. Recent surveys of both Galactic, and extra-Galactic clouds show that this is no longer the case. Hughes \emph{et al.} (2010 \& 2013b) and Meidt \emph{et al.} (2013) have attributed the observed variations in physical properties, including variations in the size-linewidth coefficient for clouds in the {\small M51, M33} and the {\small LMC} to variations in the ambient environment i.e., variation in the magnitude of interstellar pressure in the disk of a galaxy. In a related work Field \emph{et al.} (2011) argued that the Heyer \emph{et al.} data could be reconciled if clouds were to obey the Virial equilibrium modified by the external pressure. Furthermore, they argued that a cloud could possibly experience a range of pressures between 10$^{4}$ K cm$^{-3}$ - 10$^{7}$ K cm$^{-3}$. Similarly, Dobbs et al. (2011) reported that most of the clouds that formed in their simulated galactic disk were unbound. In fact, Clark \emph{et al.} (2005) demonstrated that individual star-forming clouds need not be Virially bound. \\ \\
In view of these recent suggestions about the possible dependence of physical properties of clouds on their ambient environment, here we examined this question numerically. To this end we developed hydrodynamic simulations to study formation of clouds via head-on collisions of cylindrical gas-flows having initially uniform density. We developed 11 realisations spanning a range of external pressures between 10$^{3.5}$ K cm$^{-3}$ - 10$^{8}$ K cm$^{-3}$, commensurate with that reported for different regions of the Galaxy - from the outermost to the innermost radius (e.g. Kasparova \& Zasov 2008, Ao \emph{et al.} 2013). In general we found that clouds experiencing relatively small magnitudes of external pressure, typically lower than $\sim 10^{4}$ K cm$^{-3}$ and therefore likely to be found in the outermost regions of the Galactic disk are unlikely to be of much interest from the perspective of star-formation, despite the small magnitude of gas velocity dispersion; see e.g. Figs. 3 and 4. However, a more definite statement can be made only after developing realisations that include details of the relevant atomic/molecular chemistry which presently has been substituted with a relatively simple cooling curve for the {\small ISM}.
From the respective plots shown in Figs. 3 and 4 it can also be readily inferred that the magnitude of velocity dispersion steadily increases with increasing magnitude of external pressure, though this rise is rather drastic for $P_{ext}/k_{B}\gtrsim 10^{6}$ K cm$^{-3}$. In general, the velocity dispersion, $(\sigma_{gas}/[\mathrm{km/s}]) \propto ((P_{ext}/k_{B})/[\mathrm{K \ cm^{-3}}])^{0.23}$, which is roughly consistent with the analytic prediction made by Elmegreen (1989). This is also qualitatively consistent with the trend recently reported for Galactic clouds (e.g. Rice \emph{et al.} 2016), and for extra-Galactic clouds \textbf{(e.g. Hughes \emph{et al.} 2013b)}.\\ \\
Now, although there is a dearth of literary evidence comparing the {\small N-PDF}s of clouds as a function of the magnitude of interstellar pressure, $P_{ext}$, across a wide range of environments, a few studies have reported physical properties of the dense cloud {\small G0.253+0.016}, better known as the \emph{Brick}, located close to the Galactic centre in the Central Molecular Zone ({\small CMZ}). Although this cloud has an average density $\sim10^{4}$ cm$^{-3}$, its {\small N-PDF} does not exhibit a power-law tail at the high-density end (e.g. Rathborne \emph{et al.} 2014). As is well-known, the power-law tail is a good  proxy for potentially star-forming gas in a cloud (see e.g. Kainulainen \emph{et al.}  2009; Lombardi \emph{et al.} 2015). Plots of {\small N-PDF}s shown in Figs. 5 and 6 reveal a trend where-in clouds experiencing a relatively small magnitude of external pressure, typically $P_{ext}/k_{B}\lesssim 10^{4}$ K cm$^{-3}$, tend to have a lognormal distribution and develop a distinct power-law tail at the high-density end for higher magnitudes of external pressure, $P_{ext}/k_{B}\gtrsim 10^{5}$ K cm$^{-3}$. In fact, this power-law tail often appears to be composed of two components. For extreme magnitudes of pressure typically upwards of $10^{7}$ K cm$^{-3}$, however, the {\small N-PDF} shows further steepening at the high density end which is consistent with that reported for the \emph{Brick}. This reinforces the inference that in spite of the relatively large volume density of the assembled cloud, the fraction of putative star-forming gas in such clouds is considerably small.\\ \\
These simulations also demonstrate the evolution of {\small N-PDF}s for different magnitudes of external pressure, $P_{ext}$, from a lognormal distribution to one where a power-law begins to appear at higher densities. The timescale of evolution, however, appears to depend on the magnitude of $P_{ext}$, or equivalently, on the magnitude of the inflow velocity, $V_{inf}$. We also observe that the power-law at the high-density end becomes shallower with the slope in the range (-1, -0.6), for increasing magnitude of $P_{ext}$, but less than $\sim 10^{7}$ K cm$^{-3}$. Fortunately, there is some corroborative evidence from recent studies of different regions of the Orion A cloud. Stutz \& Kainulainen (2015) reported variations in the slope of the high-density end of the {\small N-PDF} for different regions of this cloud. This slope was shallow ($\sim$-0.9), for the region having the highest concentration of {\small YSO}s. This is also the region that apparently is most affected by feedback from the nearby population of {\small O-}stars. By contrast, regions with little star-formation activity and therefore with fewer {\small YSO}s had significantly steeper power-law tails with a slope in excess of $\sim -2$. Similar conclusions were also deduced by Pokhrel \emph{et al.} (2016) in their study of the different regions of the {\small GMC} Mon R2. In fact, these authors also report that the tail at the high-density end of the {\small N-PDF} for some regions shows a break so that the tail could be fitted with two power-laws. This is also true of the {\small N-PDF}s deduced for some clouds simulated in this work as is evident from the plots for realisations 4-6 shown in Fig. 5.
The correlation between star-formation activity and a shallower tail of the {\small N-PDF} appears universal. Abreu Vicente \emph{et al.} (2015), for instance, in their extensive study of 330 {\small MC}s in the first Galactic quadrant reported similar conclusions.  The same also appears true with the {\small N-PDF}s for extra-Galactic clouds as shown by Hughes \emph{et al.} (2013a) in their study of clouds in the {\small M51}.  \\ \\
Another feature visible from the plots in Figs. 5 and 6 showing the {\small N-PDF}s is the shift in the position of the peak of a {\small N-PDF} as a function of the external pressure. For magnitudes of pressure, $P_{ext}/k_{B}\lesssim 10^{6}$ K cm$^{-3}$, this shift is not very significant, but a leftward shift towards a lower column density is readily visible in {\small N-PDF}s for higher magnitudes of pressure shown in Fig. 6. Thus, while on the one hand a stronger collision i.e., a larger magnitude of external pressure, $P_{ext}$, cycles more gas into the dense-phase, the peak of the {\small N-PDF} appears to shift towards lower column densities which is consistent with the {\small N-PDF} for the typical Galactic centre cloud such as the \emph{Brick}. Equivalently this means, a larger magnitude of external pressure causes greater segregation of gas between the dense and the rarefied phase. \emph{The upshot being that although a large magnitude of external pressure might easily cycle gas into the dense-phase (see plots shown on the upper and central-panel of Fig. 7), the fraction of gas at large column densities that can potentially form stars would diminish.} This is indeed the case as is evident from the plot on the lower-panel of Fig. 7. This latter plot shows that for intermediate magnitude of external pressure, $P_{ext}$, the fraction of cloud-mass assembled in the dense-phase steadily increased to 1\% at the time of terminating calculations. By contrast, for higher magnitudes of pressure, but $P_{ext}< 10^{7}$ K cm$^{-3}$, this fraction was as high as 3\% and thereafter, it steadily declined. In the extreme cases of $P_{ext}/k_{B}\gtrsim 10^{7}$ K cm$^{-3}$, however, it appears, gas in the shocked-slab is rapidly cycled between the dense and rarefied phases. \\ \\
These inferences are also consistent with deductions from some of the more recent surveys of Galactic clouds. Ragan \emph{et al.} (2016), for instance, using data from the {\small Hi-GAL} survey, suggest that the dense mass fraction (which they refer to as the star-forming fraction), shows a statistically significant decline with increasing Galactic radius, or equivalently, with decreasing magnitude of the interstellar pressure. This conclusion is also consistent with the one presented by Roman-Duval \emph{et al.} (2010) who had also shown a decrease in the dense gas-fraction i.e., the fraction of putative star-forming gas, with increasing Galactic radius. Furthermore, Koda \emph{et al.} (2016) also showed that the dense gas-fraction steadily declined towards the Galactic centre, although the efficiency of converting neutral gas into its dense molecular counterpart is almost 100\% at small Galactic radii where the magnitude of interstellar is significantly large, $P_{ext}/k_{B}\gtrsim 10^{7}$ K cm$^{-3}$. Results presented in the plots shown on the central and the lower-panel of Fig. 7 above are also consistent with these observational deductions.   \\ \\
Now, although the dense mass fraction shows a steady decline for the latter set of realisations at the time of terminating calculations, this fraction could increase somewhat as dense filamentary structure might start forming in the post-collapse globule as was seen to be the case in realisation 10; see upper-panel of Fig. 8. This image is interesting as it shows not only the formation of a filamentary network, but also two other prominent filaments that are likely to collide close to the centre of this globule. The foregoing discussion demonstrates that star-formation in a pristine cloud is likely to be terribly inefficient with only a small fraction of gas being cycled into the dense-phase. This conclusion is consistent with observational inferences; Murray (2011), for instance, estimated that only about 8\% of the gas in a typical {\small MW GMC} is converted into stars over its lifetime. In one of their earlier works, Fukui \& Mizuno (1991) also showed that star-formation in nearby {\small MC}s is sluggish.  \\ \\
The situation with clouds in the Galactic {\small CMZ} is even worse as a typical {\small CMZ} cloud such as the \emph{Brick} shows very little evidence of star-formation (e.g. Kauffman \emph{et al.} 2013). Although a more recent study by Marsh \emph{et al.} (2016) provides further evidence about the possible onset of star-formation in this cloud. There is, however, no immediate estimate about the fraction of gas that is possibly involved in forming stars. \emph{We note, our models in no way refute the possibility of star-formation in clouds confined by a relatively large magnitude of pressure, and in fact, do corroborate the formation of putative star-forming pockets in such clouds at a later stage of their evolution. But we also suggest that the fraction of this gas is likely to be relatively small.} \\ \\
However, our inferences contradict those reported by Bertram \emph{et al.} (2015). These latter authors reported that irrespective of the hostile nature of ambient environment, their model clouds yielded star-formation efficiency significantly higher than the rather sluggish rate reported for the \emph{Brick} by Kauffman \emph{et al.} Bertram \emph{et al.} therefore concluded that the merely unbounded nature of clouds and a stronger interstellar radiation field were not by themselves sufficient to arrest and/or inhibit star-formation in their model clouds, although increasing the unboundedness of their clouds did significantly reduce the star-formation efficiency in their simulations. However, despite this, it remained considerably higher than the one inferred for typical clouds in the Galactic {\small CMZ}. Results from their simulations led the authors to infer that star-formation in the {\small CMZ} clouds would eventually pick-up and that we are probably observing only the earliest stages of stellar-birth in these clouds. It is indeed true that star-formation activity in the \emph{Brick} is only recent (e.g. Longmore \emph{et al.} 2013, Marsh \emph{et al.} 2016), but it is also true that the {\small N-PDF} for this cloud, as discussed above, appears clipped at large column densities suggesting that star-formation will likely remain sluggish even in the future. \emph{On the contrary, our models here show that while a higher magnitude of external pressure, typically, $P_{ext}/k_{B}\gtrsim 10^{6}$ K cm$^{-3}$, undoubtedly cycles a larger fraction of gas into the dense-phase (gas-density in excess of 10$^{3}$ cm$^{-3}$), the fraction of putative star-forming gas i.e., the fraction of gas at relatively large column densities (N$_{gas}\gtrsim$ 10$^{21}$ cm$^{-2}$), however, does not show any appreciable increase. This could possibly explain why clouds close to the Galactic centre form stars sluggishly despite their relatively large volume density and richness in molecular gas.}  \\ \\
We now proceed to examine the effect of the magnitude of external pressure on the physical properties of the cloud assembled in respective realisations.  As defined in the earlier section, the gas-body comprising of particles denser  than $\sim$50 cm$^{-3}$ in a realisation was described as a cloud, the physical properties of which were calculated using Eqns. 4 - 6 above. Plots in the upper-panel of Fig. 8 made for four representative realisations show that the surface density, $\Sigma_{gas}$, of the cloud so identified at different epochs of a realisation is indeed sensitive to the magnitude of the external pressure. Not only does $\Sigma_{gas}$ for a cloud vary temporally for a given magnitude of external pressure, as can be seen from the plot on the upper left-hand panel of this figure; the plot on the upper right-hand panel of this figure shows that $\Sigma_{gas}$ also varies proportionally with $P_{ext}$. This temporal variation of $\Sigma_{gas}$ is the result of the structural changes to the post-collision slab induced by the growth of the {\small NTSI} in it. The crosses on this latter plot represent the magnitude of $\Sigma_{gas}$ for the clouds at different epochs as a function of $P_{ext}$ for these respective realisations, where-as the solid-line corresponds to $\Sigma_{gas}\propto P_{ext}$.\\ \\
 Although the plot on the right-hand panel of Fig. 8 appears to suggest that the surface density of clouds must typically increase with increasing magnitude of the interstellar pressure, we note that the trend is visible only with regard to the peak magnitude of $\Sigma_{gas}$ that the cloud acquires briefly during its evolutionary cycle. Thereafter the cloud surface-density diminished rapidly. Ragan \emph{et al.} (2016) also argued that the surface density of mass in dense clumps falls-off at small Galactic radii. This finding can be reconciled with for example, the results from simulations 10 and 11 presented in this work. In these latter realisations the surface density, $\Sigma_{gas}$, of the assembled cloud did in fact show a decline shortly after having acquired its peak magnitude as the cloud was attended to by the {\small NTSI}. \emph{This leads to the suggestion that the clouds detected at small Galactic radii may well have had larger surface densities at an earlier epoch, soon after they were assembled and presently are at a relatively advanced stage of their evolution.}  \\ \\
Evidently, the observed variation in the magnitude of $\Sigma_{gas}$ for the simulated clouds in this work is inconsistent with one of the principle conclusions of Larson (1981), that was also reinforced by observational findings reported by, for e.g. Solomon \emph{et al.} (1987) and Bolatto \emph{et al.} (2008), that clouds have an approximately constant surface density, typically $\sim$100 M$_{\odot}$ pc$^{-2}$. On the contrary, more recent surveys of Galactic clouds show that $\Sigma_{gas}$ is indeed sensitive to their ambient environment i.e., to the magnitude of interstellar pressure, $P_{ext}$, experienced by them (see review by Heyer \& Dame 2015). This is also true of extra-Galactic clouds (e.g. Hughes \emph{et al.} 2013b). The inference from these respective data being that clouds experiencing a higher magnitude of external pressure would have a larger surface density. On the other hand, if indeed the clouds obeyed the {\small SVE} then, $\Sigma_{gas}\propto P_{ext}^{1/2}$, and furthermore, if the magnitude of interstellar pressure, $P_{ext}$, in the disk of a galaxy were to be uniform then clouds would naturally be objects having a roughly uniform surface density (e.g. Elmegreen 1989). In such eventuality the Larson's third law would continue to hold. But we now know that the magnitude of pressure in the Galactic disk and indeed in the disks of other {\small MW}-like galaxies is not uniform and that physical properties of clouds are sensitive to the magnitude of interstellar pressure. \emph{The plot shown in Fig. 9, the surface density-size relation for the clouds assembled in these realisations reinforces our conclusion that clouds are unlikely to be entities having uniform surface density.} Contrary to earlier suggestions based on a number of surveys discussed above, we observe here that the surface density-size relationship for clouds is significantly steeper (than Larson's), and in fact, $\Sigma_{gas}\propto L^{-1.3}$. Similarly, a number of authors have argued that the inference that clouds have an approximately uniform surface density is likely an artefact of the specific choice of the molecular tracer (the CO) used to identify clouds (e.g. Scalo 1990, Ballesteros-Paredes \& MacLow 2002, Feldman \emph{et al.} 2012). \\ \\ \\
Next, the plots in respectively the left and right-hand panels of Fig. 10 show the temporal variation in the magnitude of internal pressure, $P_{int}/k_{B}$, in four of our representative realisations and the magnitude of internal pressure, $P_{int}/k_{B}$, as a function of the corresponding external pressure, $P_{ext}/k_{B}$, at different epochs of cloud evolution in these realisations. We remind the reader that for any given realisation the magnitude of $P_{ext}/k_{B}$ remains unchanged over the entire course of a realisation. As with the plots of the gas surface-density, the magnitude of internal pressure within a cloud also changes over the course of its evolution. It attains a peak soon after the slab is assembled and thereafter it cools rapidly, though in the process the growth of the {\small NTSI} induces a velocity dispersion within the layers of the shocked-slab. Then subsequently, as this instability grows into saturation and the velocity dispersion in the slab steadily decreases, the slab collapses to form an elongated globule within which dense filaments start forming as is visible in the lower-panel of Fig. 8 for realisation 10. The strongly self-gravitating gas that is forming filaments causes localised thermal heating which manifests itself in the form of increasing magnitude of $P_{int}$ in realisations 8 and 10, as is also visible in the characteristics shown on the left-hand panel of Fig. 10. 
However, the nature of this variation in the magnitude is such that clouds, even at relatively large magnitudes of external pressure, $P_{ext}/k_{B}\gtrsim 10^{7}$ K cm$^{-3}$, do not ever become over-pressured.  \\ \\
At best, the clouds in these realisations only briefly acquired a configuration where $P_{int}/k_{B}\sim  P_{ext}/k_{B}$ i.e., one that was closer to the {\small PVE}, before becoming pressure-confined. The timescale on which a cloud evolved was shorter the higher the magnitude of $P_{ext}$. This is interesting especially in view of the traditional belief that clouds experiencing higher magnitudes of external pressure and therefore  located preferably in the inner regions of the Galactic disk, must be gravitationally bound, as against those located in outer regions of the disk which must be pressure-confined. \emph{On the contrary, we observe that clouds at higher pressure do in fact become pressure-confined at later stages of their evolution which is dominated by a strong growth of the {\small NTSI}.} Unfortunately we cannot independently corroborate this inference observationally since the results reported by for instance, Hughes \emph{et al.} (2010, 2013b) for their survey of clouds ({\small MAGMA} and {\small PAWS}) only cover the range of interstellar pressure up to $\sim 10^{6}$ K cm$^{-3}$. Although our inference is consistent with the dynamic state of the \emph{Brick} for which the magnitude of the internal pressure, $P_{int}/k_{B}\sim 10^{8}$ K cm$^{-3}$, using the  physical properties deduced by Rathborne \emph{et al.} (2014), which is about an order of magnitude smaller than the estimated magnitude of external pressure, $P_{ext}/k_{B}\sim 10^{9}$ K cm$^{-3}$. \\ \\
\emph{Thus in view of the observed behaviour of various physical properties of clouds in this work as a function of $P_{ext}$, a simple bi-modal classification of clouds as either gravitationally bound or pressure-confined appears far-fetched.} We have seen, the dynamical state of a cloud varies temporally and the extent of variation appears to depend on the magnitude of external pressure it experiences. It may therefore be appropriate to argue that clouds rather display a spectrum of properties depending on their ambient conditions (see also, Rosolowsky 2007). An important consequence of the sensitivity of various cloud properties to the magnitude of ambient pressure, $P_{ext}$, is the corresponding variation in the size-linewidth relation. Recent work by Rice \emph{et al.} (2016) and indeed that by Heyer \emph{et al.} (2009) has conclusively demonstrated the variation in the 
coefficient of the size-linewidth relation for clouds varies as a function of their respective position in the Galactic disk. While the former authors showed that the said coefficient was significantly larger, implying a higher velocity-dispersion, for clouds in the inner Galactic disk, Heyer \emph{et al.} (2009) showed, it varied proportionally with the surface density of clouds which as we know, is itself sensitive to the magnitude of interstellar pressure. \\ \\
Variations in the coefficient of the size-linewidth relation have also been reported for molecular structures identified in the {\small LMC}, {\small M33} and {\small M51} (e.g. Hughes \emph{et al.} 2013b, Meidt \emph{et al.} 2013), which the authors attributed to variations in the local environment. This is also broadly consistent with the conclusions reported by Dobbs \& Bonnell (2007), who observed considerable steepening of the size-linewidth relation for dense structures in their numerically simulated disk. We observe a similar trend in Fig. 11 which is a plot showing the variation of the square of the size-linewidth coefficient for a cloud against its surface-density. \emph{In general, we observe that the maximum magnitude of size-linewidth coefficient increases with the increasing magnitude of $P_{ext}/k_{B}$.  Furthermore, this plot also demonstrates that each cloud evolves differently  depending on its ambient environment i.e., the magnitude of external pressure experienced by it.} This conclusion is equivalent to the interpretation of this plot (i.e., Fig. 11), by Field \emph{et al.} (2011) that observed cloud properties could likely be explained by not one unique magnitude of pressure, but a range between 10$^{4}$ K cm$^{-3}$ and 10$^{7}$ K cm$^{-3}$.
\subsection{Limitations of this work}
The simple assumption of cylindrical flows having initially uniform density and colliding head-on is ideal since turbulent flows in a galaxy are more likely to be fractal. Also, such flows are more likely to interact at a certain impact-factor. Carrol-Nellenback \emph{et. al.} (2014), for instance, have demonstrated that the slab resulting from a collision between uniform flows fragments rapidly due to the onset of various hydrodynamic instabilities. Consequently, such realisations typically showed a relatively lower core-formation rate in comparison with those in which fractal flows were allowed to collide. Similarly, the former genre of realisations showed a greater proclivity to form low-mass cores. In spite of this, conclusions presented in this work are unlikely to be altered significantly even if fractal flows were used, for the shocked slab in either case must evolve in a mutually similar fashion. However, the effect of the magnetic field towards stabilising the post-collision slab against various dynamic instabilities and the subsequent effect on the various physical properties investigated in this work must be examined. Even in this latter case we do not foresee any qualitative change to the inferences deduced in this work.

\section{Conclusions}
Numerical simulations reported in this work demonstrate that assembled clouds primarily evolve via an interplay between the {\small NTSI} and self-gravity. The {\small NTSI}  grows rapidly and its growth is more vigorous in cases where the inflow velocity is relatively large, or equivalently, the inflows are highly supersonic. Over the course of its growth, the {\small NTSI} induces significant structural changes to the post-collision cloud. Consequently, physical properties of clouds and indeed those of the gas that constitutes these clouds show significant variation with time and as a function of the magnitude of external pressure, or in other words, the magnitude of the inflow velocity. \\ \\
We observed that clouds in low-pressure environment show a greater propensity towards remaining diffuse and so, are not of much interest from the perspective of star-formation, in spite of their low gas velocity-dispersion. In general, we observe that the magnitude of  velocity-dispersion, $\sigma_{gas}$, induced within the assembled clouds in this exercise varies as $(\frac{P_{ext}/k_{B}}{[\mathrm{K\ cm}^{-3}]})^{0.23}$, which is roughly consistent with the hypothesis of clouds following energy-equipartition as against the {\small SVE} or indeed the {\small PVE}. \\ \\
While a larger fraction of gas is cycled into the cold, dense phase with an increasing magnitude of external pressure, for magnitudes of pressure in excess of $\sim 10^{6}$ K cm$^{-3}$, the fraction of potentially star-forming gas, usually characterised by a column density in excess of $\sim 10^{21}$ cm$^{-2}$, was hardly ever seen to exceed $\sim 3$\% and in fact, this fraction showed a steady decline as the assembled clouds evolved. On the other hand, in the clouds confined by an intermediate range of pressure between $\sim 10^{5}$ K cm$^{-3}$ - $\sim 10^{6}$ K cm$^{-3}$, the fraction of this putative star-forming gas showed a steady rise as the clouds evolved. This suggests, clouds of the latter type are likely to exhibit greater propensity towards forming stars. This is interesting as it implies, clouds in high-pressure environments such as those close to the Galactic centre, in spite of their relatively large average volume density, must be sluggish in forming stars as is indeed the case with the well-known cloud, \emph{Brick}, in the Galactic {\small CMZ}. \\ \\
An interesting diagnostic of potentially star-forming clouds is the appearance of the power-law tail in the {\small N-PDF}. We observe that clouds experiencing pressure magnitude, $P_{ext}/k_{B} \lesssim 10^{4}$ K cm$^{-3}$, are likely to remain diffuse so that their {\small N-PDF} always remains lognormal, or a combination of several lognormal distributions. For intermediate magnitude of pressure, $10^{5}\lesssim P_{ext}/k_{B}\lesssim 10^{6}$ K cm$^{-3}$, on the other hand, the {\small N-PDF}s steadily evolve from a purely lognormal form to one with a power-law tail at the high-density end. Furthermore, there is a distinct trend where-in the slope of this power-law tail becomes shallower (slope in the range -1 to -0.9), with increasing magnitude of external pressure, $P_{ext}$. Also, in some cases we observe, this power-law tail in fact, has a break and is composed of two constituents. Finally, for magnitudes of pressure $\gtrsim 10^{7}$ K cm$^{-3}$, however, the power-law tail shows considerable steepening. \\ \\
Furthermore, the outcome from simulations reported in this work also demonstrates that over their course of evolution, clouds evolve from a state where they obey the pressure-modified Virial equilibrium ({\small PVE}) to one where they could become pressure-confined; the latter is especially true for clouds in high-pressure environments at a magnitude of pressure upwards of $\sim 10^{7}$ K cm$^{-3}$. This appears consistent with the recently deduced physical properties for the \emph{Brick}. Finally, these simulations reconcile the observationally reported variation in the size-linewidth coefficient. Indeed, we find that its magnitude increases with an increasing magnitude of external pressure which is consistent with various recent observational inferences. In a follow-up work we will further investigate the impact of Galactic shear on cloud-properties and the effect of ambient environment on the ability of dense gas to form pre-stellar cores and eventually stars. 

\section*{Acknowledgements}
{\small SVA} is supported by the grant {\small YSS/2014/000304} of the Department of Science \& Technology, Government Of India, under the Young Scientist Scheme. This work would not have been possible without a generous grant of the Royal Astronomical Society awarded October 2015. SVA wishes to gratefully thank the Institute of Astronomy \& Astrophysics, T$\ddot{\mathrm{u}}$bingen, and the Sternwarte, Ludwig Maximilians Universit$\ddot{\mathrm{a}}$t, for extending their warm hospitality. Simulations reported in this work were developed on the bwGRiD computing cluster, a member of the German D-Grid initiative, funded by the Ministry for Education and Research (Bundesministerium f$\ddot{\mathrm{u}}$r Bildung und Forschung) and the Ministry for Science, Research and Arts Baden-W$\ddot{\mathrm{u}}$rttemberg (Ministerium f$\ddot{\mathrm{u}}$r Wissenschaft, Forschung und Kunst Baden-W$\ddot{\mathrm{u}}$rttemberg).  RK acknowledges financial support within the Emmy Noether research group on "Accretion Flows and Feedback in Realistic Models of Massive Star-formation" funded by the German Research Foundation under grant no. KU 2849/3-1. 
%
%=============================================================================

\bsp

\label{lastpage}

\end{document}